\documentclass[12pt]{report}

\usepackage[a4paper,left=3.5cm,right=2.5cm,top=2.5cm,bottom=2.5cm]{geometry} 
\usepackage{amsmath}
\usepackage{setspace} 
\usepackage{times} 
\usepackage{titlesec} 

\usepackage{hyperref} 
\usepackage{graphicx} 
\usepackage{float} 
\usepackage{caption} 
\usepackage{multirow}
\usepackage{url}
\usepackage{lscape}
\titleformat{\chapter}[block]{\normalfont\Large\bfseries}{\thechapter.}{1em}{}
\titleformat{\section}[block]{\normalfont\large\bfseries}{\thesection}{1em}{}

\begin{document}
\begin{titlepage}
    \centering

    \vspace*{-2cm} 
    \includegraphics[width=0.3\textwidth]{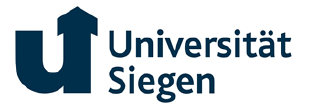} \\[1cm] 
    
    {\Large University of Siegen \\[0.5cm]
    Faculty IV: Department of Electrical Engineering and Computer Science} \\[2cm]
    
    {\Large \textbf{Master Thesis}} \\[1cm]
    
    \rule{\textwidth}{0.5pt} \\[0.3cm]
    {\large \textbf{Title: \\ Optimal Dataset Size for Recommender Systems: \\ Evaluating Algorithms' Performance via Downsampling}} \\[0.3cm]
    \rule{\textwidth}{0.5pt} \\[0.5cm]
    
    {\large Supervisors: \\[0.3cm]
    Prof. Dr. Jöran Beel \\
    Tobias Vente}\\[3cm]
    
    \begin{flushleft}
    \large
    Name: Ardalan Arabzadeh \\[0.2cm]
    Email: ardalan.arabzadeh@student.uni-siegen.de \\[0.2cm]
    Program of Study: Master of Mechatronics \\[0.2cm]
    Date of Submission: 31.01.2025
    \end{flushleft}
\end{titlepage}

\pagenumbering{roman} 
\setcounter{page}{1} 

\begin{abstract}
\thispagestyle{plain} 
This thesis investigates the potential of dataset downsampling as a strategy to optimize energy efficiency in recommender systems while maintaining competitive performance. With the increasing size of datasets posing computational and environmental challenges, this study explores the trade-offs between energy efficiency and recommendation quality in the context of Green Recommender Systems, which aim to minimize environmental impact. By applying two downsampling approaches to seven datasets, 12 algorithms, and two levels of core pruning, the research highlights that significant reductions in runtime and carbon emissions can be achieved. For instance, using a downsampling portion of 30\% in a typical case can reduce runtime by approximately 52\% compared to the full dataset, resulting in a reduction in carbon emissions of up to 51.02 \textit{KgCO$_2$e} during the training of a single algorithm on an individual dataset.

The analysis reveals that algorithm performance under different downsampling portions is influenced by factors such as dataset characteristics, algorithm complexity, and the specific downsampling configuration (scenario dependent). In particular, some algorithms, which generally showed lower absolute \textit{nDCG@10} scores compared to those that performed better, exhibited lower sensitivity to the amount of training data provided, demonstrating greater potential to achieve optimal efficiency in lower downsampling portions. For instance, on average, these algorithms retained $\sim$81\% of their full-size performance when using only 50\% of the training set. In certain configurations of the downsampling method, where the focus was on progressively involving more users while keeping the test set fixed in size, they even demonstrated higher \textit{nDCG@10} scores than when using the original full-size dataset. These findings underscore the feasibility of balancing sustainability and effectiveness, providing practical insights for designing energy-efficient recommender systems and advancing sustainable AI practices.
\\

\textbf{Keywords:} Green Recommender Systems, Energy Efficiency, Dataset Downsampling, Runtime Reduction, Carbon Emissions, Sustainability, Algorithm Performance.
\end{abstract}

\chapter*{Acknowledgements}
\setcounter{page}{2} 
\addcontentsline{toc}{chapter}{Acknowledgements} 

This thesis was conducted as part of the requirements for the Master of Mechatronics program at the University of Siegen, within the Department of Electrical Engineering and Computer Science. It was completed without any external funding, relying solely on academic and institutional resources.

Parts of the work presented in this thesis have been previously published \cite{Arabzadeh2024}, reflecting an important step in disseminating the findings to the broader academic community. Additionally, the code used for conducting the experiments is publicly available on GitHub \cite{Thesis2025github}, ensuring transparency and reproducibility of the research.

The writing and development of this thesis benefited from the use of ChatGPT for grammar and wording improvements \cite{beel2024ai}, contributing to the clarity and precision of the text. Computational resources for the experiments were partially provided by the Omni Cluster at the University of Siegen, for which I am grateful.

I would like to express my sincere gratitude to my supervisors, Professor Dr. Jöran Beel and Ph.D. student Tobias Vente, for their guidance, insightful feedback, and encouragement throughout this research. Their insights and expertise have been instrumental in shaping this work.

\tableofcontents
\listoffigures
\listoftables

\newpage
\pagenumbering{arabic}
\setcounter{page}{1}
\chapter{Introduction}
\section{Background} Recommender systems have become a cornerstone of modern digital platforms, playing a crucial role in e-Commerce, media streaming, and social networks \cite{Alfaifi2024,Sankalp2024,Schafer2001}. These systems are designed to provide personalized recommendations, improve user experiences, and drive the rapid growth of digital services \cite{Gowda2024,Maslowska2022}. Techniques like collaborative filtering, content-based filtering, and hybrid methods are commonly employed to make accurate and meaningful recommendations \cite{Schafer2007,vanMeteren2000,Thorat2015,Herlocker2000,Burke2002}. Recommender systems are commonly evaluated using metrics such as Root Mean Square Error (\textit{RMSE}), Mean Absolute Error (\textit{MAE}), Precision, Recall, and normalized Discounted Cumulative Gain (\textit{nDCG}) \cite{Cremonesi2010,Silveira2019}. Among these, \textit{nDCG} is frequently used for ranking-based evaluations, particularly \textit{nDCG@10}, which measures the relevance of the top 10 recommendations, with higher scores indicating better performance in prioritizing relevant items \cite{pmlr-v30-Wang13}.
 
Datasets play a pivotal role in the development and evaluation of recommender systems, as they define the foundation upon which algorithms operate. Publicly available datasets, such as\textit{ MovieLens} and \textit{Amazon} Reviews, provide diverse scenarios for testing recommender systems in controlled environments \cite{Alam2021,Basaran2017}. These datasets often vary in size, sparsity, and user-item interaction density, making them suitable for studying algorithm scalability, robustness, and performance under different conditions. For example, different versions of the \textit{MovieLens} datasets (\textit{100K}, \textit{1M}, \textit{10M}, \textit{20M}) are widely recognized for their high data density, well-balanced structure, and consistent rating distribution, making them a common choice for recommender systems research and experiments \cite{Harper2015}.

Datasets are essential not only for evaluating the performance of recommender systems but also for tackling challenges related to scalability and algorithmic efficiency \cite{Kuzelewska2020,Singh2020,Roy2021}. The size of these datasets can vary significantly, ranging from small-scale collections with thousands of interactions to large-scale datasets containing millions or even billions of entries. In the training phase, the runtime and resource demands become especially critical when developing algorithms that must operate efficiently in real-world applications, where large datasets and limited computational resources are often encountered \cite{Yang2021}.

As the reliance on large datasets and computational resources continues to grow in the development of recommender systems, it is important to consider their broader environmental and societal impacts. AI systems, including recommender algorithms, often require substantial computational resources, which can lead to increased energy consumption and contribute to greenhouse gas emissions \cite{sundberg2024tackling}. As climate change becomes an urgent global challenge, it is essential to recognize and address the environmental footprint of these technologies \cite{vente2024clicks}. Developing methods to measure, monitor, and minimize their carbon emissions is crucial to ensure that advancements in AI contribute to sustainable progress rather than exacerbating ecological concerns \cite{budennyy2022eco2ai}.

\section{Research Problem} The increasing reliance on recommender systems has introduced significant computational challenges, particularly in terms of scalability and efficiency. As the volume of available data increases exponentially, processing and analyzing these datasets in real time demands vast computational resources and more advanced algorithms \cite{al-jarrah2015efficient,chen2020deep}.

In general, it is observed that the performance of deep learning and AI-based algorithms, including recommender systems, improves with larger training datasets. These expansive datasets allow models to capture more intricate patterns, leading to greater accuracy and enhanced performance \cite{Sun2017,Ng2020}. However, handling massive datasets comes at a cost: substantial computational resources are required, resulting in prolonged training times, higher operational costs, and increased energy consumption \cite{al-jarrah2015efficient,Yang2021,chen2020deep}. Furthermore, the environmental impact of processing these large datasets is substantial, with significant carbon emissions associated with computational demands \cite{vente2024clicks,spillo2023sustainability,Spillo2024}. For instance, the total \textit{CO$_2$e} emissions generated by conducting all the full-paper experiments presented at ACM RecSys 2023 were equivalent to the emissions produced by 384 flights between New York (USA) and Melbourne (Australia) \cite{Lannelongue2021}.

The disparity in energy consumption between datasets of different sizes further illustrates this issue. For example, experiments using the \textit{DGCF} algorithm revealed that the energy consumed when processing the \textit{Yelp-2018} dataset was up to 1,444 times greater than that required for the \textit{Hetrec-LastFM} dataset, despite the former being only approximately 63 times larger in size \cite{vente2024clicks}. This highlights an important insight: If experiments are conducted on smaller datasets, a significant amount of energy could be saved. For example, datasets like \textit{MovieLens 10M} are commonly used for training recommender systems \cite{Harper2015}, but it is unclear whether utilizing the full 10 million instances is always necessary, especially when employing simple baseline algorithms. This raises the possibility that smaller subsets of data could achieve comparable performance for certain models while significantly reducing runtime and energy consumption. Addressing this aspect is critical to achieving more sustainable and efficient recommender systems.

With growing concerns about environmental sustainability, the focus on the environmental impact of computational systems, including recommender systems, has intensified. Addressing this challenge is the concept of Green Recommender Systems, which aim to mitigate the environmental costs of these systems \cite{vente2024clicks,Spillo2024,Arabzadeh2024}. As defined by Beel et al.\cite{Beel2024e}:\\
“\textit{Green Recommender Systems” are recommender systems designed to
minimize their environmental impact throughout their life cycle, from research
and design to implementation and operation. Green Recommender
Systems typically aim to match the performance of traditional systems but
may also accept trade-offs in accuracy or other metrics to prioritize sustainability.
Minimizing environmental impact typically but not necessarily
means minimizing energy consumption and CO$_2$ emissions.} \cite{Beel2024e}\\

\section{Research Question} As mentioned earlier, the increasing size and complexity of the datasets used in the recommender systems pose significant computational and environmental challenges. One potential solution to mitigate these challenges is downsampling the dataset. Could reducing it, perhaps to just 10\% of its original size, be sufficient to maintain performance while potentially saving up to 90\% of the energy required for computation? This raises a broader question:\\
\textit{Is it possible to achieve an acceptable trade-off between energy efficiency and performance in recommender algorithms by reducing dataset size?} \\
This study is situated within the broader context of " Green Recommender Systems", which emphasizes minimizing environmental impact while maintaining effective functionality.

\section{Research Goal} The goal of this research is to investigate the potential of dataset downsampling as a means of achieving energy-efficient recommender systems without significantly compromising performance. By exploring the relationship between dataset size and performance, this work aims to develop guidelines to minimize the environmental impact of recommender systems, making them more sustainable, possibly without or with minimum sacrificing of accuracy of the recommendation. \cite{Arabzadeh2024}.
To achieve this goal, our research objectives are as follows:

\begin{itemize}
    \item \textbf{Assess impact on recommendation quality:} This objective aims to evaluate how and to what extent different downsampling approaches affect critical performance metrics, such as \textit{nDCG@10}. By evaluating these approaches, we seek to analyze their influence on recommendation quality across varying levels of downsampling and identify thresholds beyond which performance degrades significantly.
\end{itemize}

\begin{itemize}
    \item \textbf{Quantify computational savings:} This objective involves measuring the reductions in runtime and the associated \textit{CO$_2$e} emissions achieved through downsampling of the data set across various algorithms of the recommender system. By analyzing these reductions, we aim to estimate the computational and energy savings provided by the proposed downsampling approaches, offering insights into their efficiency and environmental benefits.
\end{itemize}

\begin{itemize}
    \item \textbf{Develop practical guidelines:} Based on our findings, we will propose evidence-based recommendations on the minimal dataset size necessary for effective training and evaluation of recommender systems, with a focus on balancing performance and energy efficiency.
\end{itemize}
In this research, We hypothesize that by downsampling datasets used in recommender systems, we can reduce computational time, energy consumption, and \textit{CO$_2$e} emissions, while achieving nearly the same recommendation quality as when using full datasets\cite{Arabzadeh2024}.

\section{Contribution}
This thesis advances the field of Green Recommender Systems by investigating the interplay between dataset size, algorithmic performance, and environmental sustainability. The study demonstrates that downsampling datasets can yield substantial energy and carbon savings, while the extent of performance compromise depends on the algorithm, dataset, and downsampling method used. As a reference case, downsampling the training set to 50\% reduces runtime by $\sim$27\% to $\sim$39\% compared to training on the full-size dataset, depending on the downsampling method used. These runtime reductions correspond to significant environmental benefits, translating to carbon emission savings of approximately 26.49 \textit{KgCO$_2$e} to 38.26 \textit{KgCO$_2$e} for training a single algorithm on a single dataset.

On average, across all algorithms and datasets examined, in the specific case of 50\% downsampling, performance measured by \textit{nDCG@10} can decrease by only $\sim$36\% compared to the full dataset size or, in some cases, increase by as much as $\sim$10\%, depending on the configurations of the downsampling method and the primary focus it seeks to address. These findings emphasize the potential of optimizing dataset sizes to achieve environmentally sustainable recommender systems while maintaining competitive levels of performance. By quantifying the trade-offs between computational efficiency and algorithmic effectiveness, this work provides actionable insights into designing energy-efficient systems without significant performance degradation, offering a practical path toward sustainability in AI and Recommender systems practices.

\chapter{Related Work}
Several studies in the field of recommender systems have explored the relationship between data set size and algorithm performance efficiency, a key focus of this research. Among these, the study by Spillo et al., published shortly before the results of this thesis were reported, investigates the potential of data reduction strategies to improve sustainability in recommender systems \cite{Spillo2024}. Their analysis, based on two datasets and nine state-of-the-art algorithms, demonstrated that data reduction can significantly lower the carbon footprint of recommender systems, generally following a linear trend. While data reduction led to an expected decrease in accuracy, they also identified that the trade-off between accuracy and emissions is highly scenario-dependent, requiring careful consideration of application-specific needs. While their findings are insightful, the study's focus on a single library and a limited dataset variety restricts its generalizability. This thesis addresses these gaps by exploring a wider range of datasets, applying different levels of core pruning as preprocessing steps, analyzing two distinct downsampling approaches, and incorporating diverse algorithms, including deep learning models and more traditional algorithms such as \textit{UserKNN} and \textit{ItemKNN}, which are commonly used in research. Furthermore, these analyses leverage three widely used libraries, RecBole, LensKit, and RecPack, to provide a more comprehensive evaluation of the impact of data reduction on performance and energy efficiency.

In addition to addressing these limitations, other studies have explored related trade-offs between algorithm performance and data constraints. Notably, Bentzer and Thulin explored the trade-off between accuracy and computational efficiency in collaborative filtering algorithms under constrained data scenarios \cite{bentzer2023}. Their findings revealed that the \textit{IBCF} (Item-Based Collaborating Filtering) algorithm demonstrates superior accuracy in smaller datasets, whereas the \textit{SVD} (Singular Value Decomposition) algorithm outperforms \textit{IBCF} in terms of speed and scalability when applied to larger datasets. However, their study is limited to comparing only these two algorithms and does not provide a broader analysis of how other algorithms respond to similar constraints. To address this limitation, this thesis evaluates a broader range of algorithms, focusing on optimizing both energy efficiency and performance under varying data sizes.

Similarly, Jain and Jindal reviewed the role of sampling and filtering techniques in enhancing the computational efficiency of recommender systems \cite{jain2023sampling}. Their study provides a theoretical overview of various sampling methods discussed in different papers and, based on their evaluation, suggests specific sampling and filtering methods that are more effective for various applications and domains within recommender systems, including movies, product recommendations, and disease diagnosis systems. These methods aim to capture essential attributes of dataset by selecting a subset of data, thereby increasing the efficiency and optimizing the learning process. While their theoretical insights suggest improvements in speed and accuracy, their lack of empirical evaluation leaves a gap in understanding the practical implications of these methods. This thesis complements their theoretical work by offering empirical analyses that bridge the gap between theory and application, focusing on the practical implications of dataset reduction.

Despite extensive research on data reduction techniques in domains such as automated machine learning, artificial intelligence, and computer vision \cite{ALZOUBI2024143090,castellanos2024strategies,Castellanosgreenai,hennig2024leveragingautomlsustainabledeep,Santos2024,tornede2023towards}, their application in recommender systems remains comparatively under-explored. For instance, Zogaj et al. demonstrated that dataset downsampling in genetic programming-based AutoML systems can lead to improved computational efficiency and, in many cases, enhanced pipeline performance, especially for larger datasets. Their work highlights that optimal downsampling ratios enable exploration of more candidate pipelines, sometimes resulting in models that perform better compared to using the full dataset \cite{zogaj2021doing}. While these findings reveal the potential benefits of dataset reduction, they are not directly applicable to traditional recommender systems, where similar effects remain underexplored, particularly in the context of automated recommender systems (AutoRecSys \cite{Anand2020,Gupta2020,Vente2023a}), where vast configuration spaces need to be explored. Building on these foundational insights, this thesis adapts and applies data reduction principles specifically to recommender systems, addressing domain-specific challenges and opportunities.

\chapter{Methodology}
\section{Datasets and Preprocessing}
To conduct our experiments, we utilized seven datasets: \textit{MovieLens 100K}, \textit{MovieLens 1M}, \textit{MovieLens 10M} \cite{Harper2015}, \textit{Amazon 2018 Toys-and-Games}, \textit{Amazon 2018 CDs-and-Vinyl}, \textit{Amazon 2018 Electronics} \cite{Ni2019}, and \textit{Gowalla} \cite{Cho2011}. The key statistics for each data set before preprocessing are presented in Table \ref{tab:dataset_statistics}.

The preprocessing steps applied to the datasets include:
\begin{itemize}
    \item Removal of duplicate rows (i.e., identical user-item-rating combinations).
    \item Averaging of duplicate ratings for the same user-item pair with different rating values.
    \item Application of 10-core and 30-core pruning to retain only users and items with at least 10 and 30 interactions, respectively \cite{Lin2024,qu2024scalable}.
\end{itemize}

The statistics of the resulting dataset after preprocessing are summarized in Table \ref{tab:expanded_dataset_statistics}. It is important to note that applying 10-core pruning preserves all seven datasets for analysis. However, under 30-core pruning, only four datasets—\textit{MovieLens 100K}, \textit{MovieLens 1M}, \textit{MovieLens 10M}, and \textit{Gowalla}—contain sufficient interactions to remain viable for further evaluation. The other three datasets were excluded under this criterion due to insufficient data density. Consequently, the results and analyses in this thesis primarily utilize the 10-core pruning case with all seven datasets for consistency and comprehensiveness. However, for the specific comparison of 10-core and 30-core pruning presented in Section \ref{sec:overall_performance} (Supplementary Analysis: 10-Core vs. 30-Core Pruning), only the four datasets supporting both pruning levels are used to ensure a fair and consistent evaluation.

Additionally, two key features of the datasets, sparsity and entropy, are reported. Sparsity quantifies the proportion of missing interactions within the dataset; higher sparsity values indicate fewer observed interactions relative to all possible user-item pairs (i.e., sparsity is the complement of density). Entropy measures the diversity or uniformity of the rating distribution; higher entropy values suggest a more balanced distribution of ratings across the scale.

\begin{table}[ht]
\renewcommand{\arraystretch}{2.25} 
\centering
\caption{Statistics of the Datasets Before Preprocessing }
\label{tab:dataset_statistics}
\resizebox{\textwidth}{!}{%
\begin{tabular}{|l|l|r|r|r|r|r|r|r|}
\hline
\textbf{Dataset} & \textbf{Feedback Type} & \textbf{Users} & \textbf{Items} & \textbf{Interactions} & \textbf{Avg Int/User} & \textbf{Avg Int/Item} & \textbf{Sparsity (\%)} & \textbf{Entropy} \\ \hline
MovieLens 100K & Explicit & 943 & 1,682 & 100,000 & 106 & 59 & 93.7 & 1.46 \\ \hline
MovieLens 1M & Explicit & 6,040 & 3,706 & 1,000,209 & 165 & 269 & 95.5 & 1.45 \\ \hline
MovieLens 10M & Explicit & 69,878 & 10,677 & 10,000,054 & 143 & 936 & 98.6 & 1.33 \\ \hline
Gowalla & Implicit & 107,092 & 1,280,969 & 6,442,892 & 60 & 5 & 99.9 & - \\ \hline
Amazon Toys and Games & Explicit & 208,180 & 78,772 & 1,828,971 & 8 & 23 & 99.9 & 0.93 \\ \hline
Amazon CDs and Vinyl & Explicit & 112,395 & 73,713 & 1,443,755 & 12 & 19 & 99.9 & 1.00 \\ \hline
Amazon Electronics & Explicit & 728,719 & 160,052 & 6,739,590 & 9 & 42 & 99.9 & 1.10 \\ \hline
\end{tabular}%
}
\renewcommand{\arraystretch}{1.0} 
\end{table}

\begin{table}[ht]
\renewcommand{\arraystretch}{1.5} 
\centering
\caption{Statistics of the Datasets After Preprocessing}
\label{tab:expanded_dataset_statistics}
\resizebox{\textwidth}{!}{%
\begin{tabular}{|l|l|r|r|r|r|r|r|}
\hline
\textbf{Dataset} & \textbf{Users} & \textbf{Items} & \textbf{Interactions} & \textbf{Avg Int/User} & \textbf{Avg Int/Item} & \textbf{Sparsity (\%)} & \textbf{Entropy} \\ \hline
\multicolumn{8}{|c|}{\textbf{Preprocessed with 10-Core pruning}} \\ \hline
MovieLens 100K & 943 & 1152 & 97,953 & 103 & 85 & 90.9 & 1.46 \\ \hline
MovieLens 1M & 6,040 & 3,260 & 998,539 & 165 & 306 & 94.9 & 1.45 \\ \hline
MovieLens 10M & 69,878 & 9,708 & 9,995,471 & 143 & 1,029 & 98.5 & 1.33 \\ \hline
Gowalla & 29,858 & 40,988 & 1,027,464 & 34 & 25 & 99.9 & - \\ \hline
Amazon Toys and Games & 11,609 & 8,443 & 202,721 & 17 & 24 & 99.7 & 0.97 \\ \hline
Amazon CDs and Vinyl & 21,450 & 18,398 & 527,503 & 24 & 28 & 99.8 & 1.07 \\ \hline
Amazon Electronics & 135,867 & 49,160 & 2,260,696 & 16 & 45 & 99.9 & 1.05 \\ \hline
\multicolumn{8}{|c|}{\textbf{Preprocessed with 30-Core pruning}} \\ \hline
MovieLens 100K  & 720 & 795 & 86,295 & 119 & 108 & 84.9 & 1.44 \\ \hline
MovieLens 1M  & 5,278 & 2,830 & 971,992 & 184 & 343 & 93.4 & 1.45 \\ \hline
MovieLens 10M   & 57,496 & 8,282 & 9,671,552 & 168 & 1167 & 97.9 & 1.33 \\ \hline
Gowalla  & 1,011 & 899 & 61,429 & 60 & 68 & 93.2 & - \\ \hline
\end{tabular}%
}
\renewcommand{\arraystretch}{1.0} 
\end{table}

\section{Algorithms and Evaluation}
To conduct our experiments, we selected 12 algorithms from three widely used recommender system libraries: LensKit \cite{ekstrand2020lenskit}, RecPack \cite{recpack2022}, and RecBole \cite{xu2023towards}. Table \ref{tab:algorithms_libraries} lists the algorithms alongside their corresponding libraries. The \textit{Random} algorithm in our experiments serves as a baseline for comparison, providing a reference point to validate the performance of other algorithms. However, it is excluded from the detailed results and statistics discussed in the following sections. Additionally, two algorithms, \textit{ItemKNN} and \textit{Popularity}, are implemented using two distinct libraries, LensKit and RecPack. This duplication allows us to verify the accuracy of results, as the outputs of these implementations should ideally produce comparable performance. It also provides an opportunity to explore subtle performance differences between the same algorithms deployed from different libraries. Moreover, The \textit{Bias} algorithm is excluded from experiments on the \textit{Gowalla} dataset due to the implicit feedback nature of the data set, as this algorithm requires explicit ratings to operate effectively.

To evaluate performance, we employ the normalized Discounted Cumulative Gain at rank 10 (\textit{nDCG@10}), which measures the quality of top 10 ranking predictions \cite{pmlr-v30-Wang13}. We ensure consistent calculation logic across all libraries to facilitate a fair comparison of algorithm results \cite{schmidt2024evaluating}.

While achieving the highest possible performance for each algorithm is not the primary focus of this thesis, we performed limited grid search hyperparameter tuning to optimize key parameters \cite{alibrahim2021hyperparameter,anggoro2021performance}. Furthermore, to enhance the reliability of results, all experiments were repeated five times using different random seed values (21, 42, 63, 84, and 105), and the final results represent the average performance across these repetitions.

\begin{table}[ht]
\renewcommand{\arraystretch}{3} 
\centering
\caption{Algorithms and Corresponding Libraries}
\label{tab:algorithms_libraries}
\resizebox{\textwidth}{!}{%
\begin{tabular}{|l|l|l|l|l|l|l|l|l|l|l|l|l|}
\hline
\textbf{Algorithms} & \textbf{UserKNN} & \textbf{ItemKNN} & \textbf{Popular} & \textbf{Random} & \textbf{Bias} & \textbf{FunkSVD} & \textbf{BiasedMF} & \textbf{NeuMF} & \textbf{Popularity} & \textbf{ItemKNN} & \textbf{SVD} & \textbf{NMF} \\ \hline
\textbf{Library} & \textbf{LensKit} & \textbf{LensKit} & \textbf{LensKit} & \textbf{LensKit} & \textbf{LensKit} & \textbf{LensKit} & \textbf{LensKit} & \textbf{RecBole} & \textbf{RecPack} & \textbf{RecPack} & \textbf{RecPack} & \textbf{RecPack} \\ \hline
\end{tabular}%
}
\renewcommand{\arraystretch}{1.0} 
\end{table}

\section{Data Splitting and Downsampling Strategies}
In all experiments conducted in this thesis, the dataset is split into three subsets: 80\% of the total interactions are allocated to the training set, 10\% to the validation set, and 10\% to the test set. The validation set is utilized for hyperparameter tuning, while the test set is used for performance evaluation.

The training set is further downsampled into various portions, ranging from 10\% to 100\% (e.g., 10\%, 20\%, 30\%, and so on, up to 100\%). Importantly, the sizes of the validation and test sets remain constant across all downsampling levels. For clarity, the downsampling percentages (e.g., 10\%, 20\%, etc.) refer to the proportion of the original training set that is retained after downsampling. For instance, 10\% downsampling corresponds to 10\% of the 80\% interactions initially allocated to the training set, whereas 100\% downsampling corresponds to using the entire training set.

In this thesis, we consistently refer to downsampling in terms of the training set. While the exact phrase “n\% of the training set” may not always be explicitly stated in later chapters for simplicity, it should be understood in this context.

We employed two distinct downsampling strategies in our experiments \cite{meng2020exploring}, referred to as \textbf{User-Based} and \textbf{User-Subset}.

\textbf{User-Based Downsampling}:
In this approach, interactions for each user are randomly divided into the three subsets: training set, validation set, and test set, based on the specified proportions. For example, 10\% of each user's interactions are assigned to the test set, 10\% to the validation set, and the remaining interactions to the training set.

When downsampling, a specific fraction of each user’s training interactions is randomly selected to create the downsampled training set. For example, in a 50\% downsampling scenario, half of the remaining interactions of each user (excluding those assigned to the validation and test sets) are included in the downsampled training set. This approach ensures that all users involved in the original dataset (after preprocessing) remain represented in every subset of training, validation, and test set across all downsampling levels, which is a key factor distinguishing this downsampling strategy from the second approach.

\textbf{User-Subset Downsampling}:
In the second downsampling strategy, we downsample the dataset by selecting a random subset of users for each downsampling step (i.e., at higher downsampling levels, a larger portion of users is included). The interactions of these selected users are split such that 10\% of the total interactions in the original dataset (after preprocessing) are allocated to the validation set, 10\% to the test set, and the remaining interactions of the selected users are assigned to the training set.

To ensure the validation and test sets maintain a fixed number of interactions across all downsampling levels, the proportion of each user's interactions allocated to these sets decreases as more users are included at higher downsampling portions. Consequently, the size of the training set grows progressively, as desired, with increasing downsampling levels according to the expected portions (10\%, 20\%, ..., 100\%).

Unlike the User-Based approach, not all users from the original dataset are included in every downsampling step. Instead, only a selected subset of users is included at each level. As a result, an increasing proportion of users contribute to the validation and test sets as the downsampling portion grows. Despite this, the retention of a fixed size for the validation and test sets in terms of the number of interactions across all downsampling portions, as well as ensuring the consistency of having the same users across the training, validation, and test sets within each downsampling step, is preserved. Unlike the previous, more commonly used method, this setup enables an analysis of how reducing the dataset size by limiting the number of users impacts runtime, \textit{CO$_2$e} emissions, and the obtained \textit{nDCG@10} score compared to the full size of the dataset. Additionally, it simulates real-world applications where an increasing number of users is continuously involved in the dataset in online systems.

The codes and configurations for all experiments conducted in this study are openly accessible on GitHub \cite{Thesis2025github}.

\chapter{Results}
This chapter provides a comprehensive analysis of the performance and efficiency of various recommender algorithms across multiple datasets, evaluated using two distinct downsampling approaches: \textbf{User-Based} and \textbf{User-Subset} downsampling. The main focus of this chapter is twofold: firstly, analyzing the performance of algorithms in terms of their achieved \textit{nDCG@10} scores at different downsampling portions, and secondly, examining the corresponding performance-efficiency trade-offs.

In this context, two key terms are used repeatedly throughout the upcoming chapters: Normalized \textit{nDCG@10} score and Relative Performance. The Normalized \textit{nDCG@10} score, sometimes referred to as Normalized Performance, represents the absolute \textit{nDCG@10} scores obtained by algorithms, scaled to a range of 0 to 1 to allow fair comparisons between datasets when averaging results. On the other hand, Relative Performance evaluates the efficiency of downsampling by expressing the performance of each algorithm in a given downsampling portion as a percentage of its performance (measured by the obtained \textit{nDCG@10} score) in the full size of the dataset (100\%), which serves as the baseline. When the term ‘more efficient’ is used, it refers to achieving a score closer to the score obtained on the full size of the dataset. These definitions are provided to avoid ambiguity, as the terms will be used consistently throughout the following sections and chapters. These investigations consider overall trends, as well as algorithm-specific and dataset-specific behaviors, discussed in subsequent sections.

The chapter begins with an overview of overall performance trends, highlighting the general behaviors and performance averages of algorithms across all datasets. As part of this analysis, a supplementary section delves into the effects of 10-core and 30-core pruning, offering additional insights into the impact of different core pruning levels on algorithm performance.

Subsequently, a deeper analysis is conducted, focusing on algorithm-specific and dataset-specific behaviors to provide a more granular understanding of performance variations. This analysis highlights the influence of algorithmic and dataset characteristics on the observed trends and uncovers patterns of similar behavior across algorithms and datasets.

Finally, the chapter explores runtime efficiency and the associated reductions in carbon emissions and energy consumption achieved through downsampling. This section provides valuable insights into the sustainability implications of the examined downsampling approaches, further emphasizing the trade-offs between performance and efficiency.

Unless explicitly stated in the supplementary section on core pruning, all results presented in this chapter are based on the 10-core pruned versions of the datasets.

\section{Overall Performance Trends}
\label{sec:overall_performance}
\begin{figure}[H]
    \centering
    \begin{tabular}{cc}
        \includegraphics[width=1.0\linewidth]{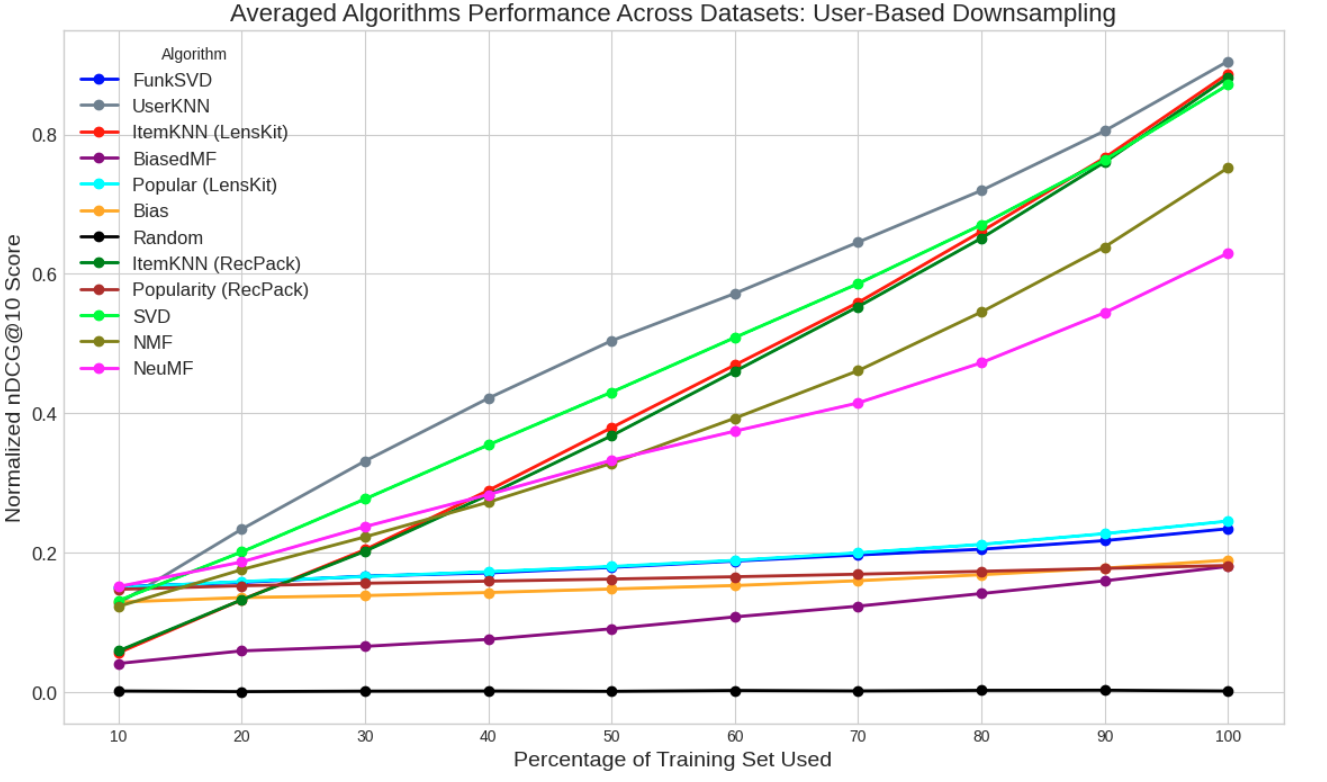} \\
        \includegraphics[width=1.0\linewidth]{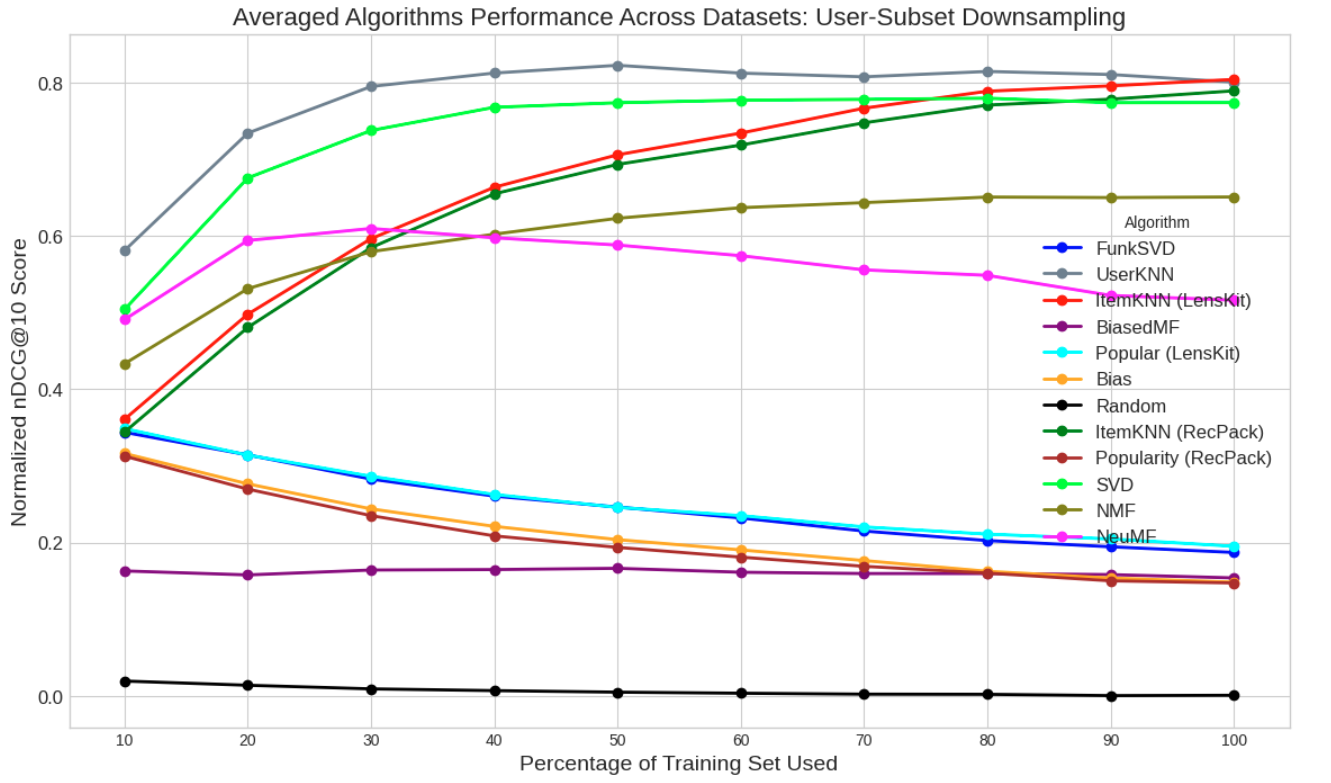}  
    \end{tabular}
    \caption{Normalized and Averaged \textit{nDCG@10} Scores of Algorithms Across All Datasets for User-Based (upper plot) and User-Subset (lower plot) Downsampling Approaches.}
    \label{fig:Averaged-Normalized}
\end{figure}

To understand the behavior of recommender algorithms under different downsampling scenarios, we first analyze their normalized \textit{nDCG@10} performance averaged across all datasets. Figure \ref{fig:Averaged-Normalized} illustrates the normalized and averaged performance trends for the User-Based and User-Subset approaches. The normalized scores of each algorithm are averaged across datasets to capture how performance evolves as the training set size increases, up to the full dataset size (100\%).

In Figure \ref{fig:Averaged-Normalized} (User-Based), we observe that the normalized \textit{nDCG@10} scores for all examined cases steadily increase as the downsampling portion grows. For example, at 30\%, 50\%, and 70\% of the training set, the average normalized score for the \textit{SVD} algorithm rises from 0.27 to 0.43 and 0.58, respectively, demonstrating a clear upward trend. In contrast, results obtained using User-Subset approach reveals a more varied pattern across different groups of algorithms. For instance, while the \textit{SVD} algorithm shows a slight upward trend, with average normalized \textit{nDCG@10} scores of 0.73, 0.77, and 0.78 at the same training set portions (30\%, 50\%, and 70\%), and appears to reach a saturation point around the 50\%, the \textit{Popular} (LensKit) algorithm exhibits a declining trend, achieving normalized scores of 0.28, 0.24, and 0.22, respectively. This divergence highlights differences in algorithm behavior under both downsampling approaches, which will be explored in greater detail in the following sections.

\begin{figure} [H]
    \centering
    \includegraphics[width=1\linewidth]{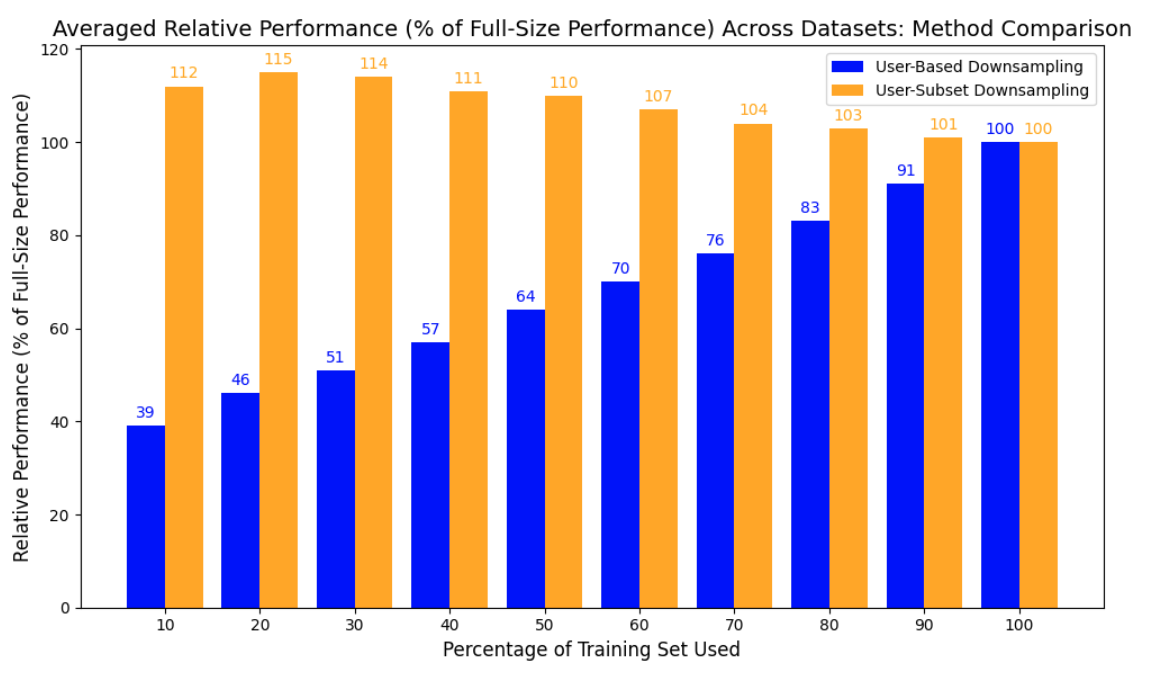}
    \caption{Averaged Relative Performance (\% of Full Dataset Performance) of All Algorithms Across All Datasets at Each Downsampling Portion for Both User-Based and User-Subset Downsampling Approaches.}
    \label{fig:Averaged-Relative}
\end{figure}

While figures \ref{fig:Averaged-Normalized} focus on normalized algorithm performance, the next step evaluates relative performance efficiency compared to full-size training data. Figure \ref{fig:Averaged-Relative} presents the averaged relative performance of all algorithms across datasets at each downsampling portion, expressed as a percentage of full dataset performance. This analysis offers insights into the efficiency trade-offs between the User-Based and User-Subset approaches. Specifically, relative performance percentages highlight each downsampling method's ability (not directly comparable between methods) to retain effectiveness as training data is reduced.

At 10\% of the training set, the User-Based approach achieves an average of $\sim$39\% of its full dataset performance, while the User-Subset approach records $\sim$112\%. For the User-Subset approach, values at this portion range from $\sim$54\% to $\sim$193\% depending on algorithm and dataset, highlighting variations influenced by specific configurations. This suggests that certain algorithms can retain performance very effectively under the User-Subset method, especially with minimal training data, as discussed in greater detail later.

Similarly, at 50\% of the training set, the User-Based approach achieves $\sim$64\% of its full dataset performance, while the User-Subset approach achieves $\sim$110\%. These findings indicate that the behavior of algorithms varies with the strategy and focus used to reduce training data, with the User-Subset approach exhibiting strong retention and, in some cases, performance in terms of \textit{nDCG@10} scores exceeding full-dataset levels at lower portions, while the User-Based approach shows an expected steady improvement as the amount of training data increases.

 The underlying potential reasons for this different behavior between the two approaches will be discussed in the Discussion and Interpretation chapter.

\subsubsection{Supplementary Analysis: 10-Core vs. 30-Core Pruning} 
This supplementary analysis explores the impact of core pruning levels on algorithm performance under both the User-Based and User-Subset downsampling approaches. Specifically, we analyze two aspects: (1) the averaged and normalized \textit{nDCG@10} scores across all algorithms and datasets and (2) the averaged relative performance (as a percentage of the \textit{nDCG@10} score obtained at full size) to assess the efficiency of each pruning level. It is important to note that this comparison is based on the four datasets (\textit{MovieLens 100K}, \textit{MovieLens 1M}, \textit{MovieLens 10M}, and \textit{Gowalla}) that support both 10-core and 30-core pruning, ensuring consistency in evaluation.

\begin{figure}[H]
    \centering
    \begin{tabular}{cc}
        \includegraphics[width=1.0\linewidth]{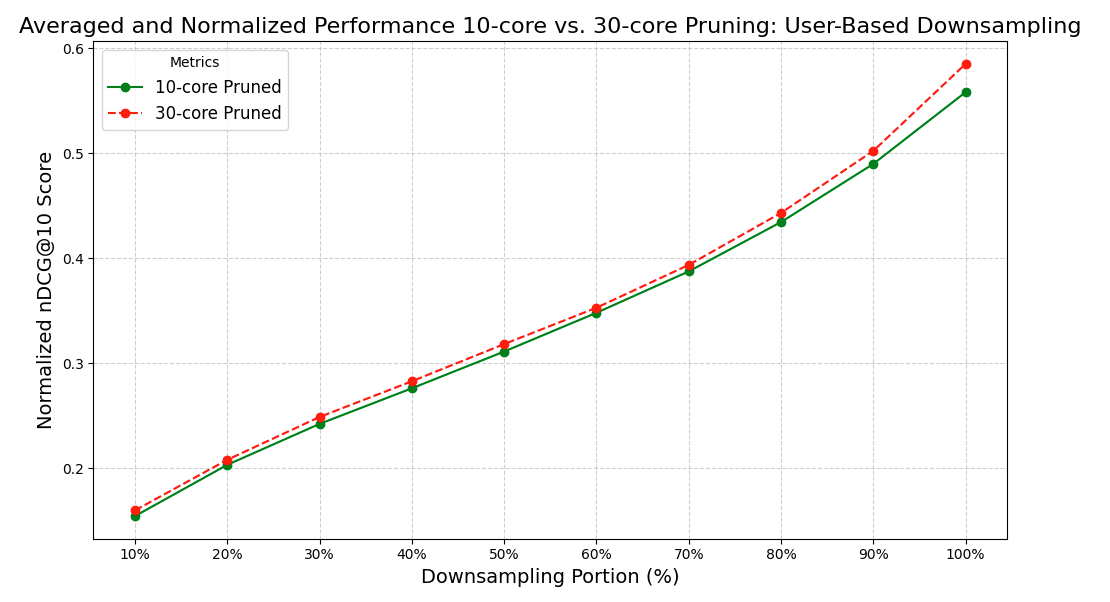} \\
        \includegraphics[width=1.0\linewidth]{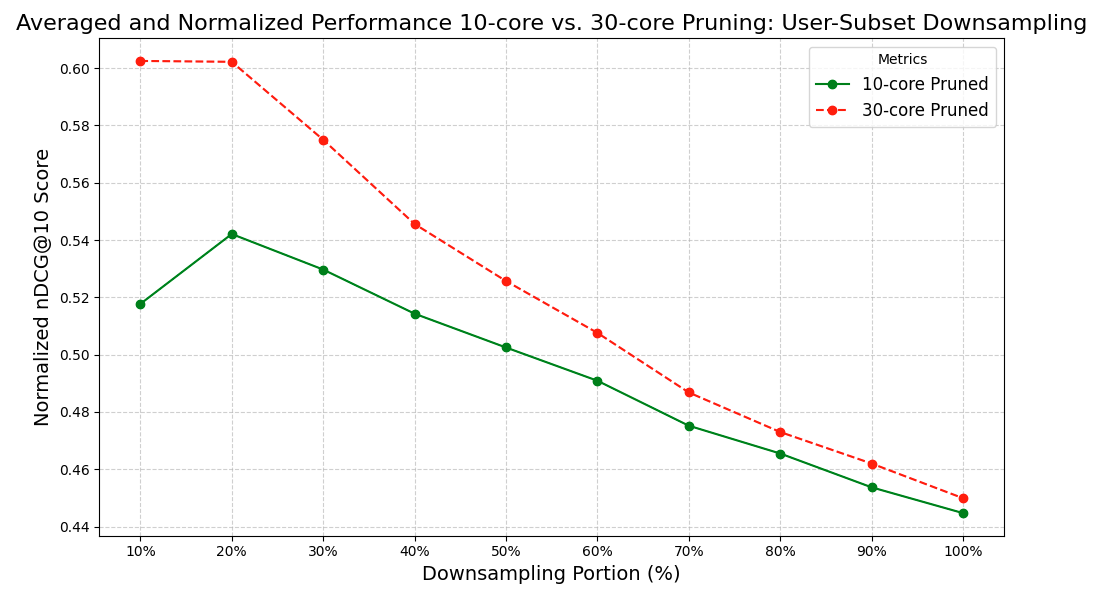}  
    \end{tabular}
    \caption{Normalized and Averaged \textit{nDCG@10} Scores Across All Algorithms and Datasets: Comparison of 10-Core vs. 30-Core Pruning for User-Based (Upper Plot) and User-Subset (Lower Plot) Downsampling Approaches.}
    \label{fig:10vs.30-Normalized}
\end{figure}

Figure \ref{fig:10vs.30-Normalized} presents the averaged and normalized \textit{nDCG@10} scores across all datasets and algorithms at each downsampling portion for the 10-core and 30-core pruning levels under the User-Based and User-Subset approaches.

In the User-Based approach (Figure \ref{fig:10vs.30-Normalized}), the results reveal that the 30-core pruning consistently outperforms 10-core pruning in all portions of the training set. For instance, at 30\% of the training set, the averaged and normalized \textit{nDCG@10} score for 10-core pruning is approximately 0.2424, while the 30-core pruning achieves a slightly higher score of 0.2490, reflecting a relative improvement of about 2.7\%. At 70\%, the scores rise to 0.3876 and 0.3938 for the 10-core and 30-core pruning levels, respectively, demonstrating an improvement of $\sim$1.6\% in this case.

A similar trend is observed in the User-Subset approach. At 30\% of the training set, the averaged and normalized \textit{nDCG@10} score for 10-core pruning is around 0.5297, compared to 0.5750 for the 30-core pruning, showing a relative improvement of 8. 5\%. At 70\%, the normalized scores increase to 0.4752 and 0.4868 for the 10-core and 30-core pruning levels, respectively, further supporting the advantage of 30-core pruning in this approach. It is important to note that the general downward trend observed for the User-Subset approach in Figure \ref{fig:10vs.30-Normalized} reflects average performance across all algorithms and datasets, with the primary purpose of broadly evaluating the effect of core pruning. As shown in Figure \ref{fig:Averaged-Normalized} and discussed in following sections, relative performance varies significantly with algorithm complexity and dataset characteristics, highlighting unique trends in specific algorithm-dataset combinations.

These results suggest that denser interaction networks in 30-core datasets provide richer training data, leading to consistently higher performance (\textit{nDCG@10} scores) under both downsampling methods. This observation aligns with findings by Beel et al.\cite{Beel2019}, which will be further discussed in the Discussion and Interpretation chapter.

\begin{figure}[H]
    \centering
    \begin{tabular}{cc}
        \includegraphics[width=1.0\linewidth]{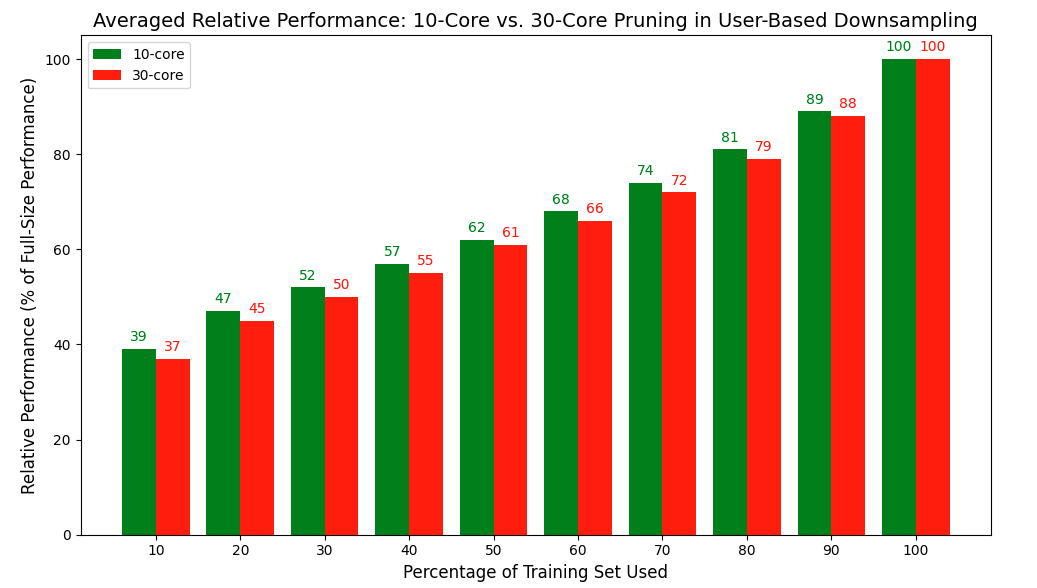} \\
        \includegraphics[width=1.0\linewidth]{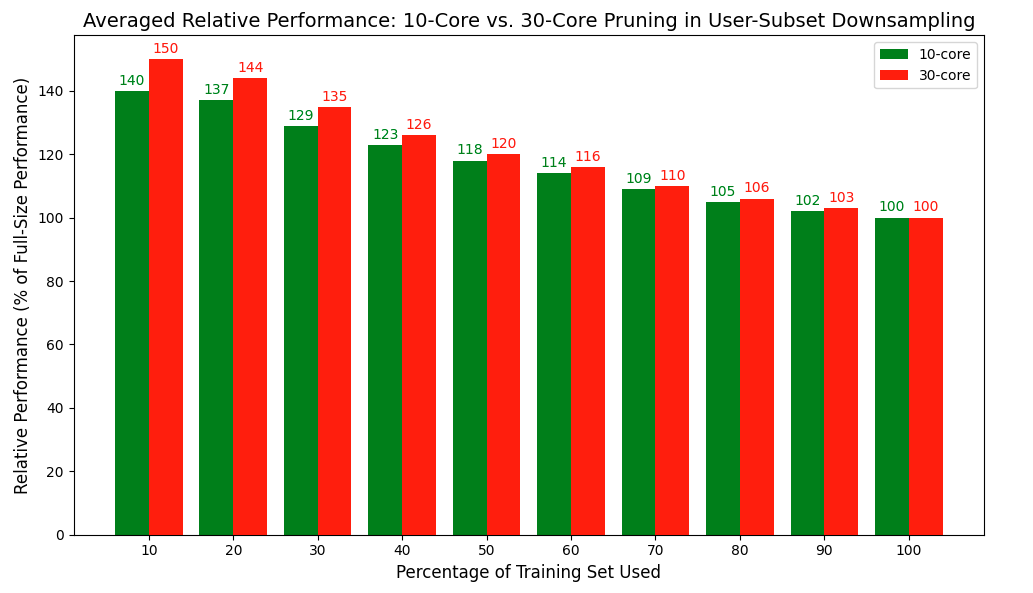}  
    \end{tabular}
    \caption{Averaged Relative Performance (\% of Full Dataset Performance) Across All Algorithms and Datasets: Comparison of 10-Core vs. 30-Core Pruning for User-Based (Upper Plot) and User-Subset (Lower Plot) Downsampling Approaches.}
    \label{fig:10vs.30-Relative}
\end{figure}

Building on this analysis, Figure \ref{fig:10vs.30-Relative} delve deeper into the efficiency of each pruning level, comparing the averaged relative performance (as a percentage of full-size performance) under both the User-Based and User-Subset approaches.

In the User-Based method (Figure \ref{fig:10vs.30-Relative}), the 10-core pruning is slightly more efficient than the 30-core pruning (i.e., the obtained \textit{nDCG@10} scores at downsampled portions are closer to the score obtained at 100\% of the dataset). For example, at 30\% of the training set, the averaged relative performance for 10-core pruning is around 52\%, compared to $\sim$50\% for the 30-core pruning. This trend persists across all downsampling portions, indicating that, despite achieving slightly lower normalized \textit{nDCG@10} scores, on average, algorithms sacrifice slightly less performance relative to their full training set size under 10-core pruning compared to 30-core pruning, demonstrating slightly greater efficiency for our data reduction purposes.

Conversely, the User-Subset plot in Figure \ref{fig:10vs.30-Relative} shows a different trend, where 30-core pruning demonstrates superior efficiency. For instance, at 30\% of the training set, the averaged relative performance of algorithms for 10-core pruning is approximately 129\%, while 30-core pruning achieves $\sim$135\%, both surpassing the full dataset performance (as discussed earlier). This shift in efficiency trend can be attributed to the fact that, as shown in Figure \ref{fig:10vs.30-Normalized} (User-Subset plot), The normalized \textit{nDCG@10} scores for algorithms under 30-core pruning improve more significantly at lower downsampling portions compared to higher portions, as evidenced by a notable gap between the 10-core and 30-core pruning versions at these lower portions. This leads to a greater relative performance gain compared to their 100\% training size performance, ultimately surpassing the 10-core pruned version in terms of efficiency. Thus, in the User-Subset approach, 30-core pruning not only enhances normalized \textit{nDCG@10} scores but also achieves better efficiency in retaining relative effectiveness than 10-core pruning.

\section{Algorithm-Specific Analysis}
\begin{figure}[H]
    \centering
    \begin{tabular}{cc}
        \includegraphics[width=1.0\linewidth]{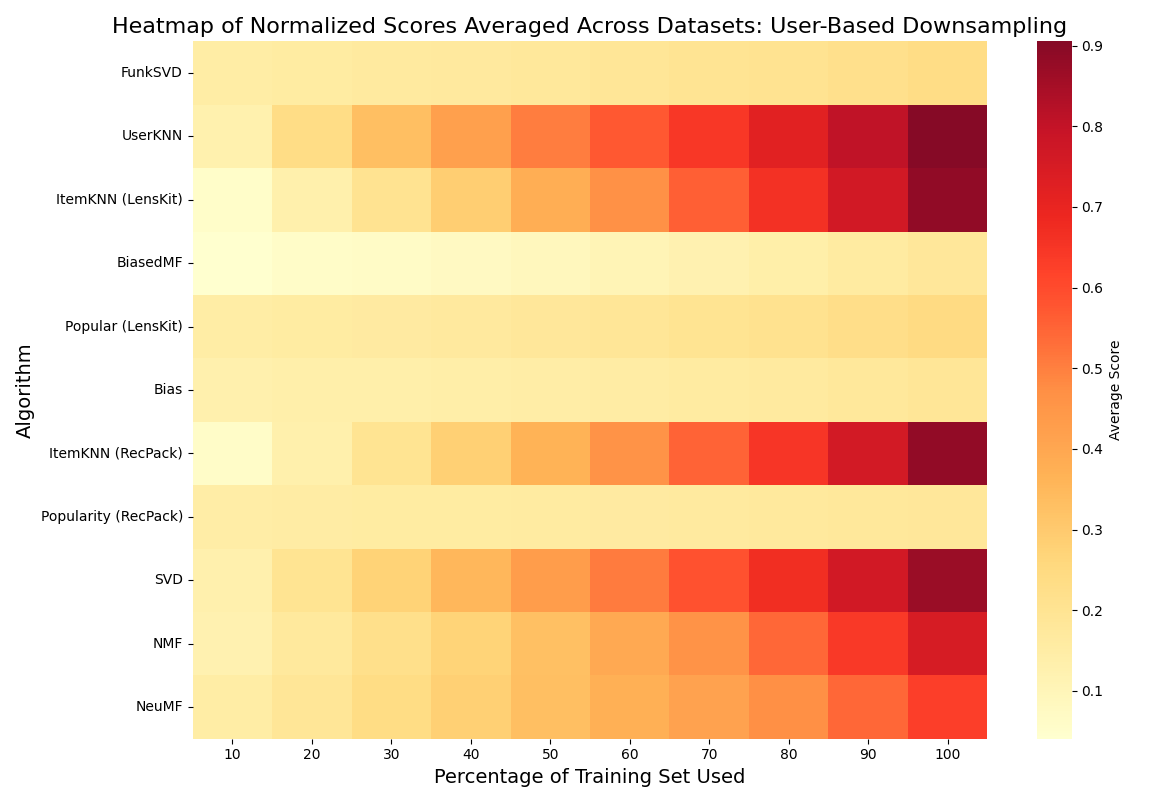} \\
        \includegraphics[width=1.0\linewidth]{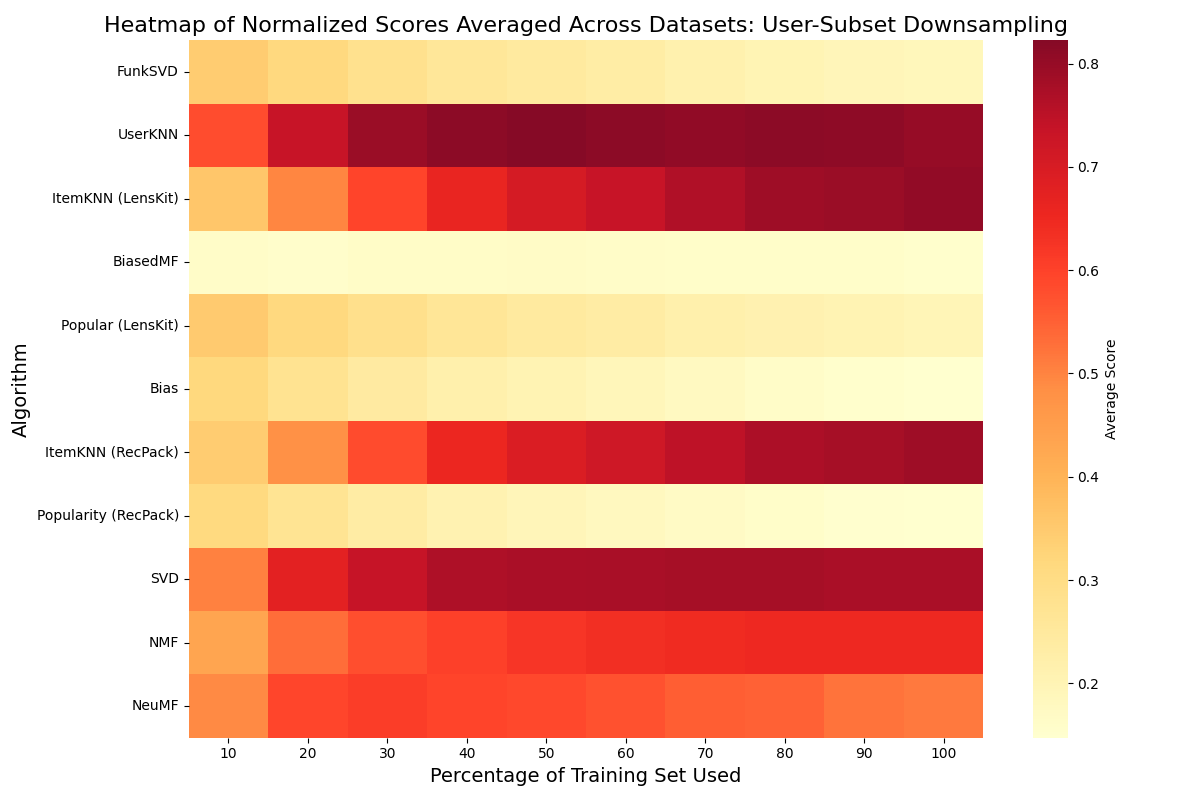}  
    \end{tabular}
    \caption{Algorithm-Specific Heat maps of Normalized Average \textit{nDCG@10} Scores Across All Datasets: User-Based Downsampling (Upper Plot) and User-Subset Downsampling (Lower Plot).}
    \label{fig:HeatMap-Algorithm}
\end{figure}

The algorithm-specific analysis of performance and efficiency delves into the distinct behaviors of individual algorithms, highlighting patterns that emerge when grouping algorithms with similar tendencies. This analysis is based on heat map visualizations that display the normalized average \textit{nDCG@10} scores for each algorithm across different downsampling portions. These heat maps, constructed for both the User-Based and User-Subset approaches, reveal two distinct groups of algorithms, as illustrated in Figure \ref{fig:HeatMap-Algorithm}.

Group 1 algorithms, comprising \textbf{\textit{UserKNN}}, \textbf{\textit{ItemKNN} (LensKit)}, \textbf{\textit{ItemKNN} (RecPack)}, \textbf{\textit{SVD}}, \textbf{\textit{NMF}}, and \textbf{\textit{NeuMF}}, distinguish themselves with significantly higher \textit{nDCG@10} scores, especially at higher downsampling portions in both approaches. They exhibit a consistent increase in normalized \textit{nDCG@10} scores as the downsampling portion increases. This upward trend is particularly pronounced in the User-Based approach. In the User-Subset approach, the trend remains, albeit with a slightly reduced gradient of improvement. This behavior is visually reflected in the heat map, with a gradient transitioning from light yellow (lower scores) to dark red (higher scores) as the downsampling portion increases.

Group 2 algorithms, including \textbf{\textit{FunkSVD}}, \textbf{\textit{Bias}}, \textbf{\textit{Popular} (LensKit)}, \textbf{\textit{Popularity} (RecPack)}, and \textbf{\textit{BiasedMF}}, display a steadier performance pattern. In the User-Based approach, their normalized scores show only a slight upward trend, shifting from light yellow to a slightly darker yellow as the downsampling portions increase. However, in the User-Subset approach, their performance slightly declines, transitioning from darker yellow at smaller portions to lighter yellow at larger portions. These results suggest that Group 2 algorithms, typically simpler models, not only achieve lower performance in terms of \textit{nDCG@10} scores compared to Group 1 but also respond differently to variations in downsampling approaches.

The patterns observed in these heat maps align closely with those in the overall trends section, where normalized performance was represented by linear graphs \ref{fig:Averaged-Normalized}. The similarity in behavior within each group underscores the feasibility of analyzing efficiency and performance trends by treating algorithms with similar characteristics as representative groups. This approach simplifies the analysis without sacrificing depth, allowing a focus on broader patterns rather than individual algorithm nuances.

\begin{figure}[H]
    \centering
    \begin{tabular}{cc}
        \includegraphics[width=1.0\linewidth]{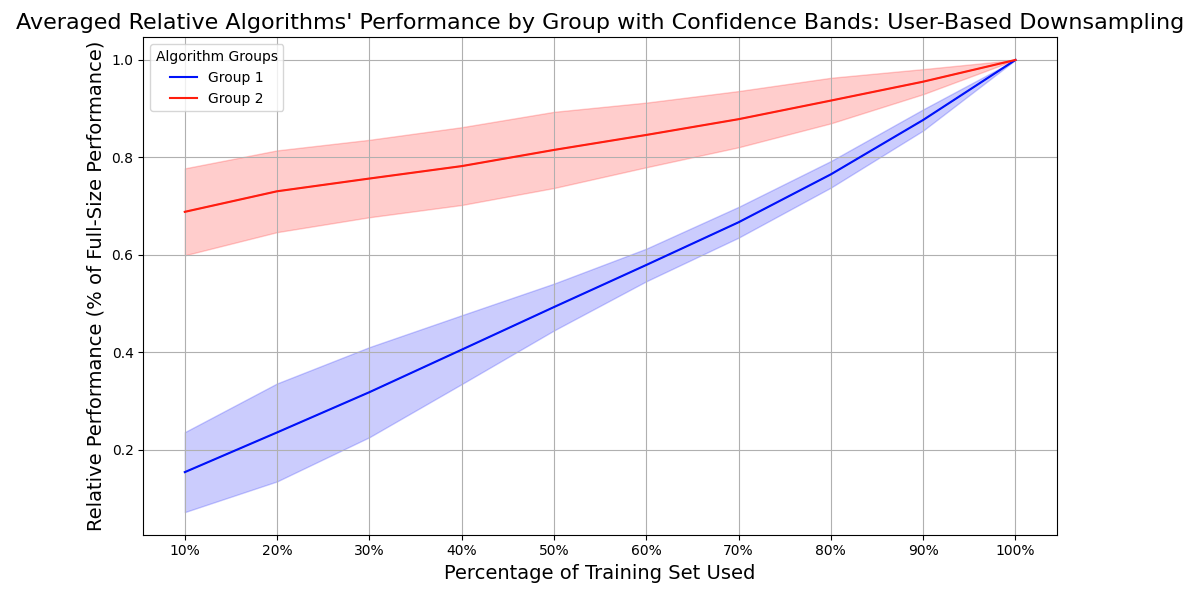} \\
        \includegraphics[width=1.0\linewidth]{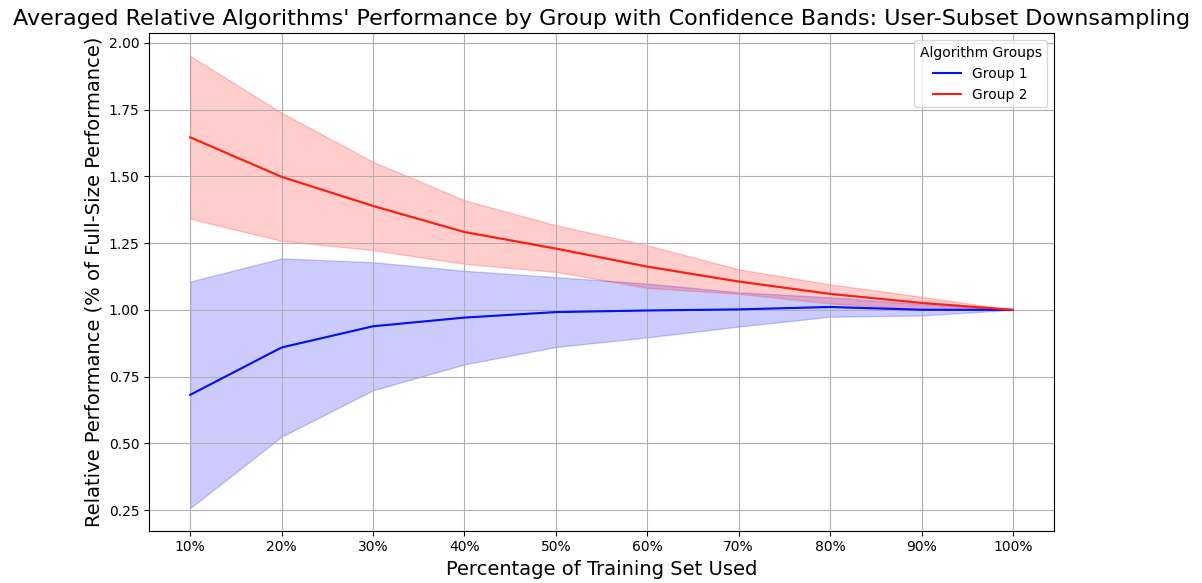}  
    \end{tabular}
    \caption{Average Relative Performance (\% of Full Dataset Baseline) for Group 1 and Group 2 Algorithms Across All Datasets: User-Based Downsampling (Upper Plot) and User-Subset Downsampling (Lower Plot).}
    \label{fig:Confidence-Algorithm}
\end{figure}

To further explore efficiency, the analysis evaluates the average relative performance of each algorithm compared to its performance at the full dataset size across all datasets, grouping algorithms into Group 1 and Group 2. Figure \ref{fig:Confidence-Algorithm} visualizes this through line graphs with confidence bands for both downsampling approaches. Preliminary observations suggest that in the User-Based approach, both groups show an upward trend as the training set size increases. For instance, Group 1 algorithms begin at approximately 15\% of their full-size performance when using 10\% of the training set and steadily climb to $\sim$76\% at 80\%. Group 2 algorithms, while starting from a higher baseline of $\sim$68\%, exhibit more gradual improvements, reaching around 91\% at 80\% training set size. This indicates that, under the User-Based approach, increasing the training data volume for a fixed set of users across all portions allows algorithms in both groups to approach their full dataset baseline performance.

However, in the User-Subset approach, a divergence arises: while Group 1 algorithms maintain a similar upward trend, Group 2 algorithms behave differently, performing better at lower downsampled portions compared to larger ones. For example, Group 2 algorithms achieve around 164\% relative performance at 10\% training set size but drop to approximately 105\% at 80\%.

The relative efficiency of algorithms is evident in the placement of lines in these charts. A higher position at a specific downsampling portion signifies that less performance is sacrificed compared to the full dataset size score as the dataset is reduced. In this context, Group 2 algorithms (generally simpler, more basic models) demonstrate higher efficiency at downsampled portions in both approaches, retaining more of their performance when trained on smaller subsets. Although this group of algorithms achieves lower absolute \textit{nDCG@10} scores than Group 1, they exhibit greater potential for energy efficiency performance trade-offs, achieving $\sim$81\% of their full-size performance with only 50\% of the training set in the User-Based approach and $\sim$123\% in the User-Subset approach.

\section{Dataset-Specific Analysis}
In this section, we shift the focus from algorithms to datasets to examine how the characteristics of the dataset influence the performance and efficiency of recommender system algorithms under varying levels of downsampling. This analysis has two main objectives: identifying common patterns among dataset groups and understanding their roles in shaping algorithm performance and efficiency.

\begin{figure}[H]
    \centering
    \begin{tabular}{cc}
        \includegraphics[width=1.0\linewidth]{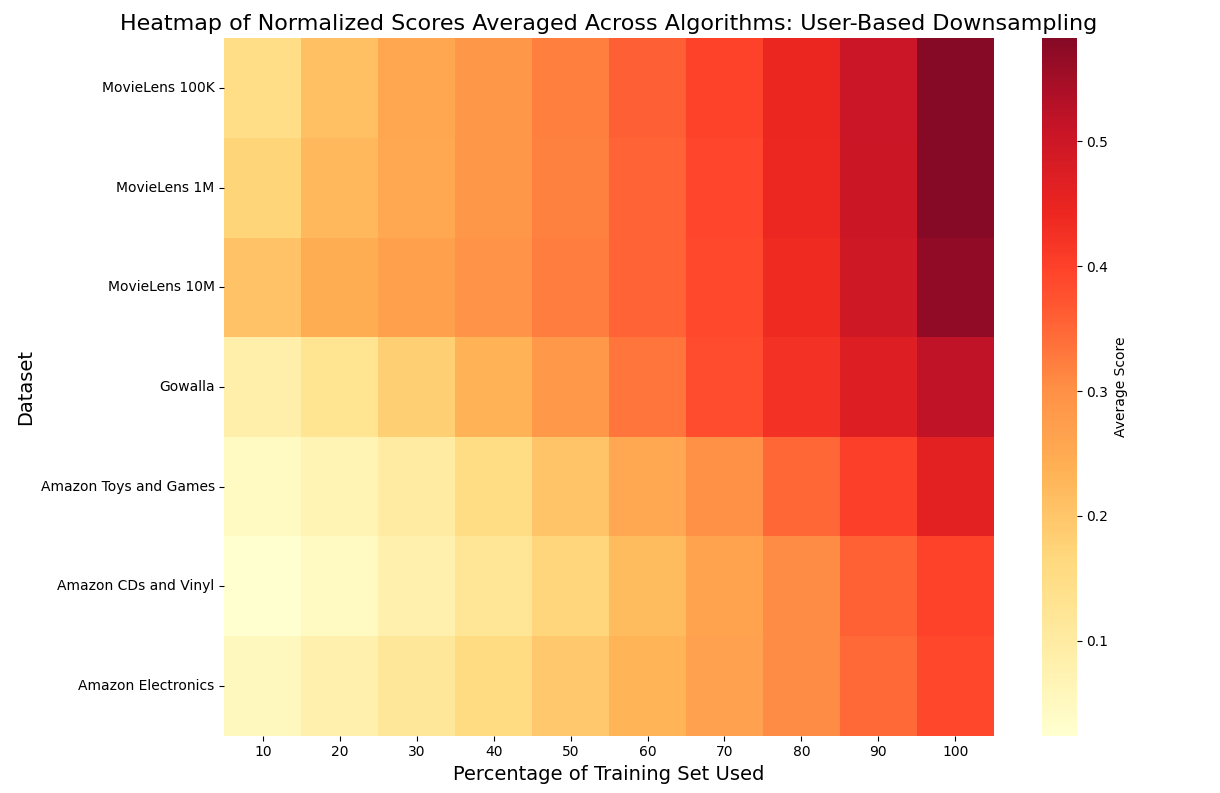} \\
        \includegraphics[width=1.0\linewidth]{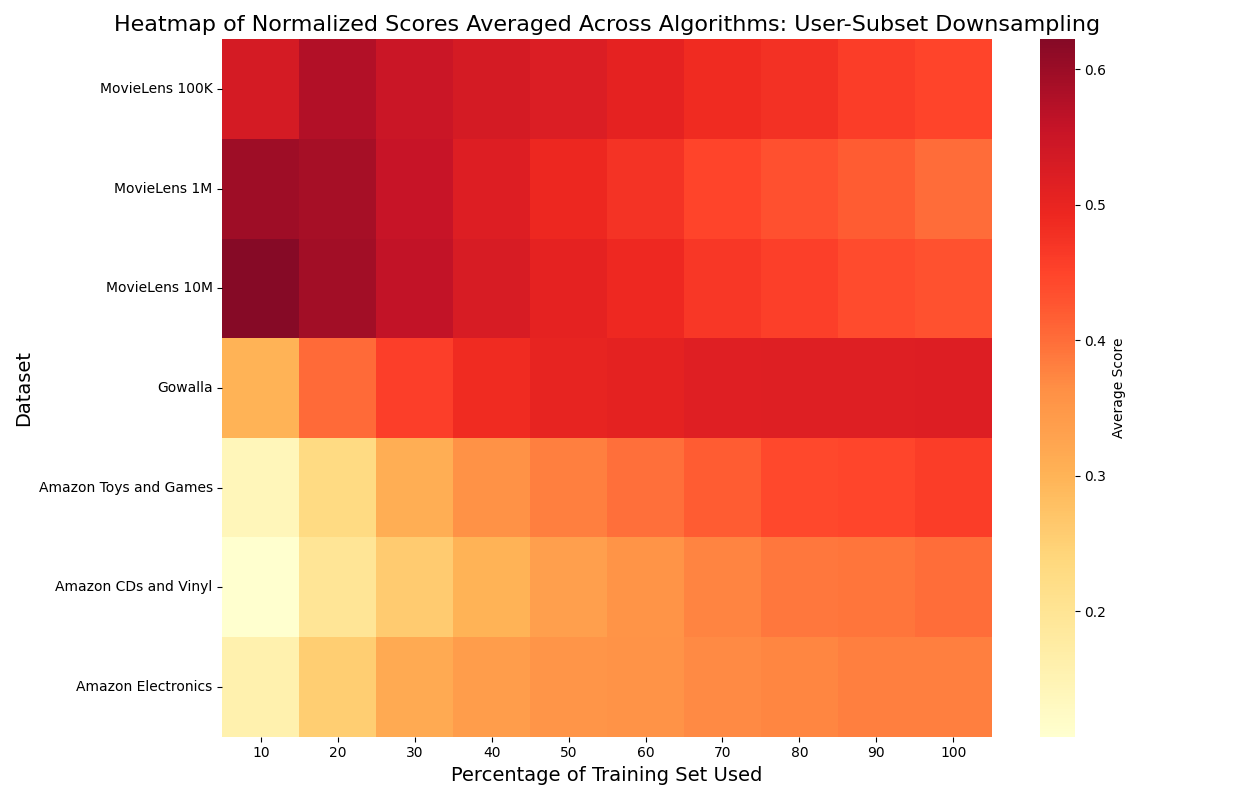}  
    \end{tabular}
    \caption{Dataset-Specific Heat maps of Normalized Average \textit{nDCG@10} Scores Across All Algorithms: User-Based Downsampling (Upper Plot) and User-Subset Downsampling (Lower Plot).}
    \label{fig:HeatMap-Dataset}
\end{figure}

To achieve this, we generated two heat maps, one for the User-Based and another for the User-Subset downsampling methods (Figure \ref{fig:HeatMap-Dataset}). Like in the previous section, these heat maps show normalized and averaged \textit{nDCG@10} scores for all algorithms across different downsampling portions, but this time, with rows representing individual datasets instead. The heat maps reveal that the datasets can be logically grouped as follows:

\textit{\textbf{MovieLens}} Group: Comprising \textbf{\textit{MovieLens 100K}}, \textbf{\textit{MovieLens 1M}}, and \textbf{\textit{MovieLens 10M}} datasets.

\textit{\textbf{Amazon}} Group: Including \textbf{\textit{Amazon Toys and Games}}, \textbf{\textit{Amazon CDs and Vinyl}}, and \textbf{\textit{Amazon Electronics}} datasets.

\textit{\textbf{Gowalla}} Group: Represented solely by the \textbf{\textit{Gowalla}} dataset, as the single implicit feedback type dataset.

This grouping is justified not only by the shared origins of the datasets within the same dataset collection, such as the \textit{Amazon} or \textit{MovieLens} collections, but also by the distinct behavioral patterns these groups exhibit in the heat maps.

In the User-Based heat map, the \textit{MovieLens} group consistently shows stronger normalized \textit{nDCG@10} scores, particularly at higher downsampling portions, represented by darker red colors. In contrast, the \textit{Amazon} group demonstrates lower performance, while \textit{Gowalla} occupies an intermediate position between the two. This pattern becomes even more pronounced in the User-Subset heat map. Here, the \textit{MovieLens} datasets exhibit a reverse trend: their normalized \textit{nDCG@10} scores decrease as the downsampling portions increase, transitioning from dark red (high scores) to light red (lower scores). Meanwhile, \textit{Amazon} datasets and \textit{Gowalla} show upward trends, with \textit{Gowalla} outperforming \textit{Amazon} in normalized scores at higher portions.

To delve deeper into overall relative performance, we calculated the relative efficiency of algorithms, defined as the percentage of their performance at each downsampling portion relative to their full-size performance. For each dataset group, these relative values were averaged across all algorithms to represent the overall efficiency of the group.

\begin{figure}[H]
    \centering
    \begin{tabular}{cc}
        \includegraphics[width=1.0\linewidth]{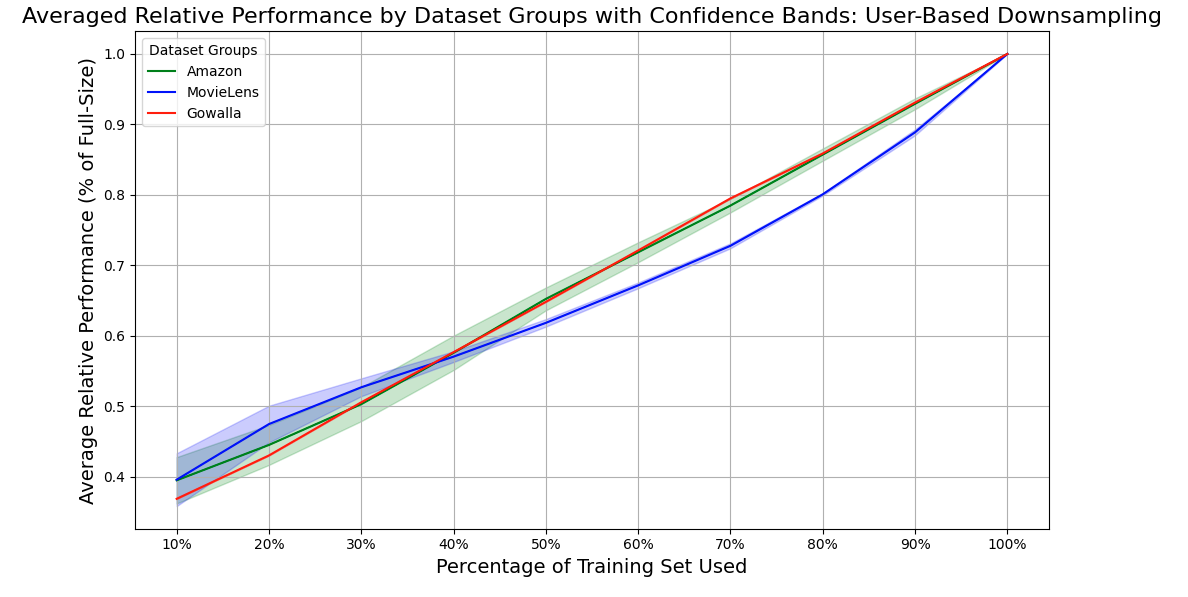} \\
        \includegraphics[width=1.0\linewidth]{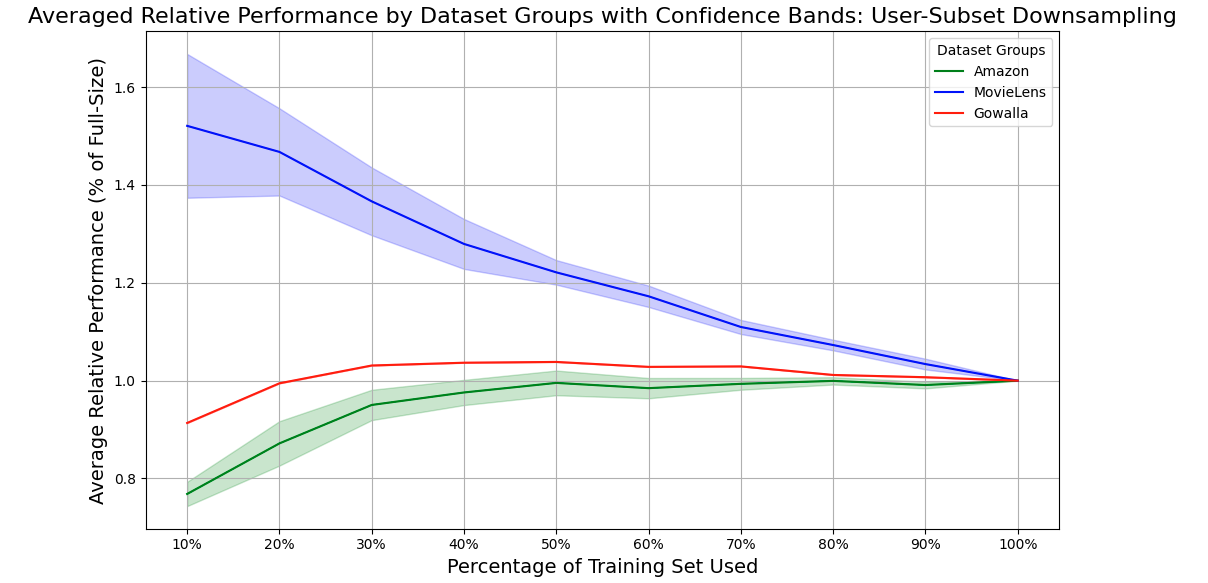}  
    \end{tabular}
    \caption{Average Relative Performance (\% of Full Dataset Baseline) for Dataset Groups Across All Algorithms: User-Based Downsampling (Upper Plot) and User-Subset Downsampling (Lower Plot).}
    \label{fig:Confidence-Dataset}
\end{figure}

Figure \ref{fig:Confidence-Dataset} illustrates these results with confidence bands, where applicable. The \textit{Gowalla} dataset, being the sole member of its group, lacks confidence bands due to the absence of variability within the group.

In the User-Based plot of Figure \ref{fig:Confidence-Dataset}, the \textit{MovieLens} group demonstrates slightly higher efficiency (averaged relative performance) than the \textit{Amazon} and \textit{Gowalla} groups up to around 40\% downsampling. For instance, at 20\% downsampling, the \textit{MovieLens} group retains an average relative performance of $\sim$47\%, compared to $\sim$44\% for \textit{Amazon} and $\sim$43\% for \textit{Gowalla}. However, beyond this point, the \textit{Amazon} and \textit{Gowalla} groups become more efficient, losing less performance relative to their full-size scores. At 70\% downsampling, \textit{Amazon} and \textit{Gowalla} retain $\sim$78\% and $\sim$79\% of their full-size performance, respectively, while \textit{MovieLens} drops to $\sim$72\%. Overall, from an efficiency perspective in downsampled dataset sizes which is a primary focus of this research, the differences between the examined groups of datasets in this downsampling approach are not substantial.

The User-Subset approach presents a different trend. The \textit{MovieLens} group maintains the highest efficiency (average relative performance) but shows a downward trend, indicating a relative drop in performance as downsampling portions increase. At 20\% downsampling, the \textit{MovieLens} group retains $\sim$146\% of its full-size performance, while \textit{Gowalla} and \textit{Amazon} retain $\sim$99\% and $\sim$87\%, respectively. At 70\% downsampling, the relative performance of \textit{Gowalla} is closely aligned with that of \textit{Amazon}, with algorithms achieving $\sim$102\% and $\sim$99\% of their full-size scores, respectively.

It is important to emphasize that the plots in Figure \ref{fig:Confidence-Dataset} reflect the average relative performance across all algorithms. For example, the upward trend observed for the \textit{Amazon} group in the User-Subset plot does not imply that all algorithms follow this trend. As noted in Figure \ref{fig:Confidence-Algorithm} of the Algorithm-Specific Analysis section, different algorithm groups can exhibit contradictory relative behaviors.

To identify the most efficient configurations with minimal performance sacrifice during downsampling, it is essential to integrate insights from both Figure \ref{fig:Confidence-Algorithm} (Algorithm-Specific Analysis) and Figure \ref{fig:Confidence-Dataset} (Dataset-Specific Analysis). This combined perspective will help to pinpoint the best combinations of algorithms and dataset groups, thereby enabling the development of energy-efficient and effective recommender systems.

\section{Runtime Efficiency and Carbon Emission Savings}
\begin{figure} [H]
    \centering
    \includegraphics[width=1\linewidth]{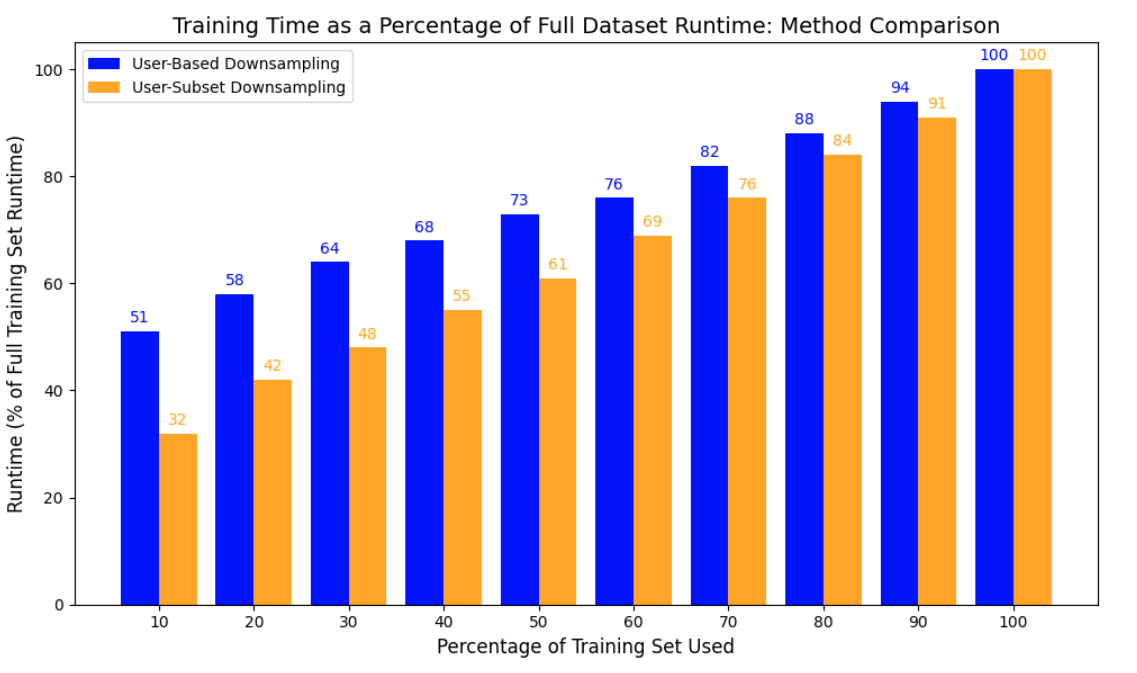}
    \caption{Runtime Reduction Analysis: User-Based vs. User-Subset Downsampling Approaches Across Various Training Set Sizes.}
    \label{fig:Runtime}
\end{figure}

To further evaluate the efficiency of the two downsampling approaches used in this investigation, we analyzed their impact on runtime reduction during algorithm training and the corresponding reduction in \textit{CO$_2$e} emissions. We calculated the average runtime across all algorithms and all seven 10-core pruned datasets for each downsampling portion. Figure~\ref{fig:Runtime} presents the average runtime percentages for both User-Based downsampling and User-Subset downsampling approaches, relative to the required runtime for the full dataset.

As illustrated, downsampling the training data using User-Based approach to 30\%, 50\%, and 70\% reduces the runtime for training and evaluation phases to $\sim$64\%, $\sim$73\%, and $\sim$82\% of the runtime required for the full dataset, respectively. 

The energy required for a single execution of a recommender algorithm on a dataset is estimated to be around 0.51 \textit{kWh} \cite{vente2024clicks}. Considering that each algorithm typically undergoes 10 hyperparameter tuning configurations, and applying the global average of 481 \textit{gCO$_2$e} per \textit{kWh} as a conversion factor \cite{ember2024carbon}, we further incorporated a scaling factor of 40 to account for additional tasks such as prototyping, debugging, reruns, and other preliminary activities \cite{vente2024clicks}. Using these values, the potential carbon equivalent emissions savings from downsampling the training set to 30\%, 50\%, and 70\% compared to the full dataset for training of a single algorithm on an individual dataset are calculated as follows:

\textbf{30\%} downsampling:
\begin{equation}
(100\% - 64\%) \times 0.51 \, \text{\textit{kWh}} \times 10 \times 481 \, \text{\textit{gCO$_2$e}/kWh} \times 40 \approx 35.32 \, \text{\textit{KgCO$_2$e}}.
\end{equation}
\textbf{50\%} downsampling:
\begin{equation}
(100\% - 73\%) \times 0.51 \, \text{\textit{kWh}} \times 10 \times 481 \, \text{\textit{gCO$_2$e/kWh}} \times 40 \approx 26.49 \, \text{\textit{KgCO$_2$e}}.
\end{equation}
\textbf{70\%} downsampling:
\begin{equation}
(100\% - 82\%) \times 0.51 \, \text{\textit{kWh}} \times 10 \times 481 \, \text{\textit{gCO$_2$e/kWh}} \times 40 \approx 17.66 \, \text{\textit{KgCO$_2$e}}.
\end{equation}

Similarly, the same analysis was conducted for the User-Subset downsampling approach.
As shown, downsampling the training data to 30\%, 50\%, and 70\% reduces the runtime for training and evaluation phases to $\sim$48\%, $\sim$61\%, and $\sim$76\% of the runtime required for the full dataset, respectively. Using the same calculation methodology, the estimated potential carbon equivalent emissions savings for the User-Subset approach are as follows:\\

\textbf{30\%} downsampling:
\begin{equation}
(100\% - 48\%) \times 0.51 \, \text{\textit{kWh}} \times 10 \times 481 \, \text{\textit{gCO$_2$e/kWh}} \times 40 \approx 51.02 \, \text{\textit{KgCO$_2$e}}.
\end{equation}
\textbf{50\%} downsampling:
\begin{equation}
(100\% - 61\%) \times 0.51 \, \text{\textit{kWh}} \times 10 \times 481 \, \text{\textit{gCO$_2$e/kWh}} \times 40 \approx 38.26 \, \text{\textit{KgCO$_2$e}}.
\end{equation}
\textbf{70\%} downsampling:
\begin{equation}
(100\% - 76\%) \times 0.51 \, \text{\textit{kWh}} \times 10 \times 481 \, \text{\textit{gCO$_2$e/kWh}} \times 40 \approx 23.54 \, \text{\textit{KgCO$_2$e}}.
\end{equation}

These calculations provide an approximate assessment of the \textit{CO$_2$e} emission reductions achieved by training a single algorithm on a single dataset, derived exclusively from the runtime reductions observed associated with downsampling. They assumed that the same hardware was utilized for both the full and the downsampled data sets and assumed a near-linear correlation between runtime, energy consumption, and carbon emissions. This assumption aligns with the principles outlined by the ML \textit{CO$_2$e} Impact calculator tool \cite{lacoste2019quantifying}.

Our analysis reveals that the User-Subset downsampling method consistently outperforms the User-Based approach in reducing the runtime during the training phase across all portions. In every point-to-point comparison of downsampling levels, the User-Subset method achieves a more significant runtime reduction. On average, across all downsampling portions from 10\% to 90\%, the User-Subset approach demonstrates an additional $\sim$16\% reduction in runtime compared to the User-Based method. This enhanced efficiency translates directly into equivalent improvements in energy savings and corresponding reductions in \textit{CO$_2$e} emissions.

\chapter{Discussion and Interpretation}
Our analysis provides valuable insights into the trade-offs between dataset size reduction and algorithmic performance in the context of energy-efficient recommender systems. One of the most striking observations is the variability in algorithm performance across different downsampling methods and datasets.

Regarding the \textbf{User-Based} downsampling method, we observe that algorithms generally exhibit an upward trend in performance with increasing downsampling portions. This indicates that providing the models with larger training sets, while keeping the number of users fixed and consistent across all portions, consistently results in improved algorithm performance. As shown in Figure \ref{fig:Confidence-Algorithm}, the sensitivity to training data reduction varies significantly across algorithms. Some algorithms experience a substantial drop in their performance, as measured by the obtained \textit{nDCG@10} scores, when compared to their full-size dataset performance. In contrast, other algorithms, typically more basic models that inherently have simpler patterns for recommendation, exhibit less sensitivity to training data reduction. These basic algorithms often achieve scores closer to their full-size performance at lower training data volumes. As a result, they can be identified as more efficient options for scenarios where absolute performance is not the primary priority, and computational or resource efficiency is emphasized.

Furthermore, regarding the effect of dataset characteristics in the User-Based approach, as illustrated in Figure \ref{fig:Confidence-Dataset}, sparser datasets, such as those of the Amazon collection, demonstrate greater efficiency in retaining algorithmic performance at higher levels of downsampling (portions above 40\%). However, the observed differences between the datasets are not substantial, suggesting that the impact of dataset sparsity in this approach is relatively minimal compared to other factors.

Overall, the consistent upward trend observed across all algorithms and datasets in this more commonly used downsampling method aligns with the broadly accepted principle in AI and machine learning that "more data leads to better performance." This trend underscores the importance of larger training data volumes for achieving optimal results in most recommendation system applications. 

However, an interesting phenomenon arises with the \textbf{User-Subset} downsampling approach, where certain algorithms outperform their full dataset performance under specific conditions. This contradictory behavior raises important questions about the underlying factors driving these differences and warrants further investigation.

A detailed analysis of algorithm performance on individual datasets reveals distinct trends for the \textbf{User-Subset} approach. Algorithms classified as poorer performing based on \textit{nDCG@10} scores (referred to as Group 2 in earlier sections) exhibit a consistent downward trend as the downsampling portion increases, irrespective of the dataset group. In contrast, algorithms with higher performance (Group 1) demonstrate varied behaviors. On \textit{Amazon} and \textit{Gowalla} datasets, these algorithms generally show an upward trend with increasing downsampling portions, while on \textit{MovieLens} datasets, they follow a downward trend similar to Group 2 algorithms. 

We hypothesize that these varying trends across datasets and algorithms stem from the \textbf{design} of the \textbf{User-Subset} downsampling method, intrinsic \textbf{dataset characteristics}, and the \textbf{nature of the algorithms} themselves. In the User-Subset approach, increasing the downsampling portion involves incorporating a larger number of users while maintaining fixed proportions of interactions (10\% of the full dataset) in the validation and test sets. This setup requires progressively adjusting the train-to-test ratio for each user's interactions as more users are added to the downsampling portion. To ensure that the total number of interactions in the validation and test sets remains constant across all downsampling portions, the proportion of each user's interactions allocated to these sets is reduced as more users are included, while the training portion is increased. This design guarantees that the validation and test sets stay consistent in size, while the training set expands incrementally as more users are included.

It is important to note that this approach introduces a fundamental difference compared to the User-Based method: the composition of the validation and test sets changes with each downsampling portion as new groups of users are included. While the amount of data reserved for validation and test sets remains statistically constant, the dynamic composition of users introduces some variability in the test and validation sets. This variability could reflect real-world scenarios where user subsets and interactions are not static. While it may contribute to variations and unexpected observed performance trends for certain algorithms, this dynamic setup also offers valuable insights into the robustness of algorithm behavior across different user populations. In addition, it introduces factors that influence performance, which will be discussed in further detail.

Despite these practical limitations, the User-Subset method is evaluated in this thesis as a case study to investigate how reducing the dataset size by decreasing the number of users impacts runtime, carbon footprint reduction, and the efficiency-performance trade-off, compared to reducing interactions while keeping the number of users fixed. In this setup, two critical factors are likely to influence performance: the change in the \textbf{number of users} included in the data subsets and the change in the \textbf{train/test ratio} of each user’s interactions within each subset.

To examine these factors, we conducted experiments with the \textit{UserKNN} algorithm, a representative of Group 1 algorithms, which shows contrasting trends on \textit{Amazon} (upward) and \textit{MovieLens} (downward) datasets. Two representative datasets, \textit{MovieLens 100K} and \textit{Amazon Toys and Games}, were selected for further analysis. For each dataset, we performed two sets of experiments: 

\begin{itemize}
    \item Fixing the number of users while varying the train/test ratio of each user's interactions across downsampling portions.
    \item Fixing the train/test ratio of each user's interactions while varying the number of users across downsampling portions.
\end{itemize}

The results indicate that when the number of users is increased while keeping the train/test ratio of each user constant, the performance of \textit{UserKNN} remains relatively stable, with only minor variations. This suggests that merely increasing the number of users without improving the quality of their interactions (e.g., denser interactions) has a limited impact on algorithm performance in terms of the obtained absolute score. As an example, when the same algorithms are applied to the full-size \textit{MovieLens 10M} dataset, they do not necessarily achieve significantly higher \textit{nDCG@10} scores compared to the full-size \textit{MovieLens 100K} dataset, despite the former having nearly 100 times more users. Both datasets share similar user interaction densities, leading to comparable performance and trends under the same train/test set ratios.

\begin{figure}[H]
    \centering
    \begin{tabular}{cc}
        \includegraphics[width=1.0\linewidth]{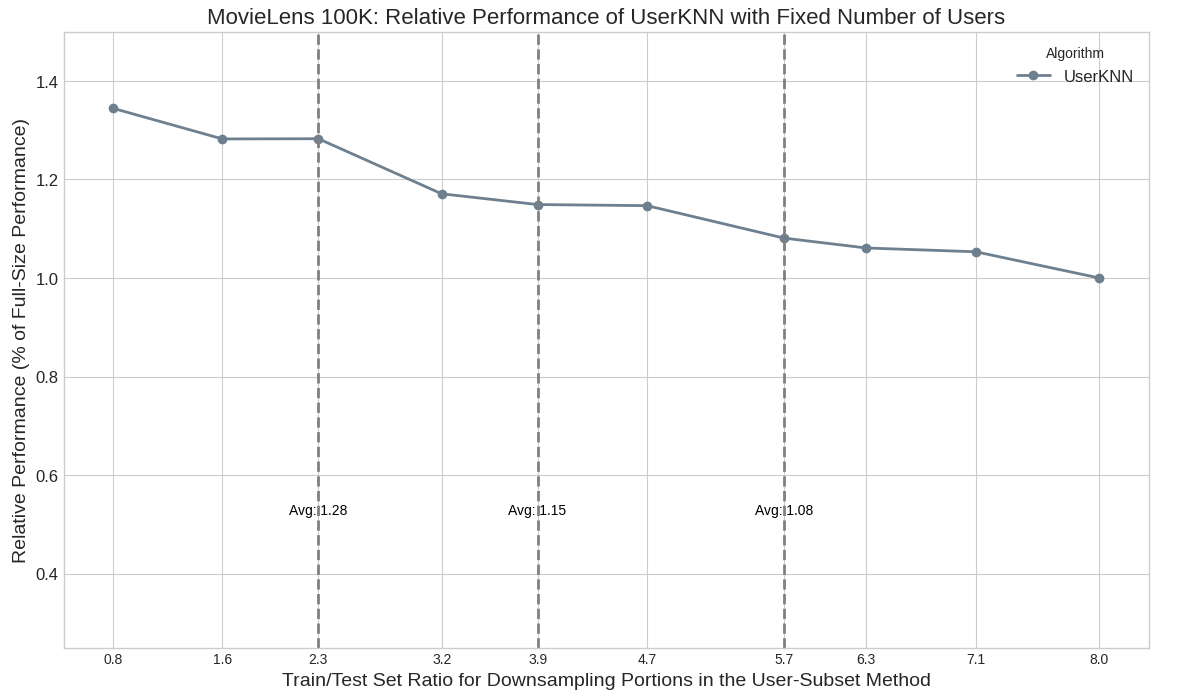} \\
        \includegraphics[width=1.0\linewidth]{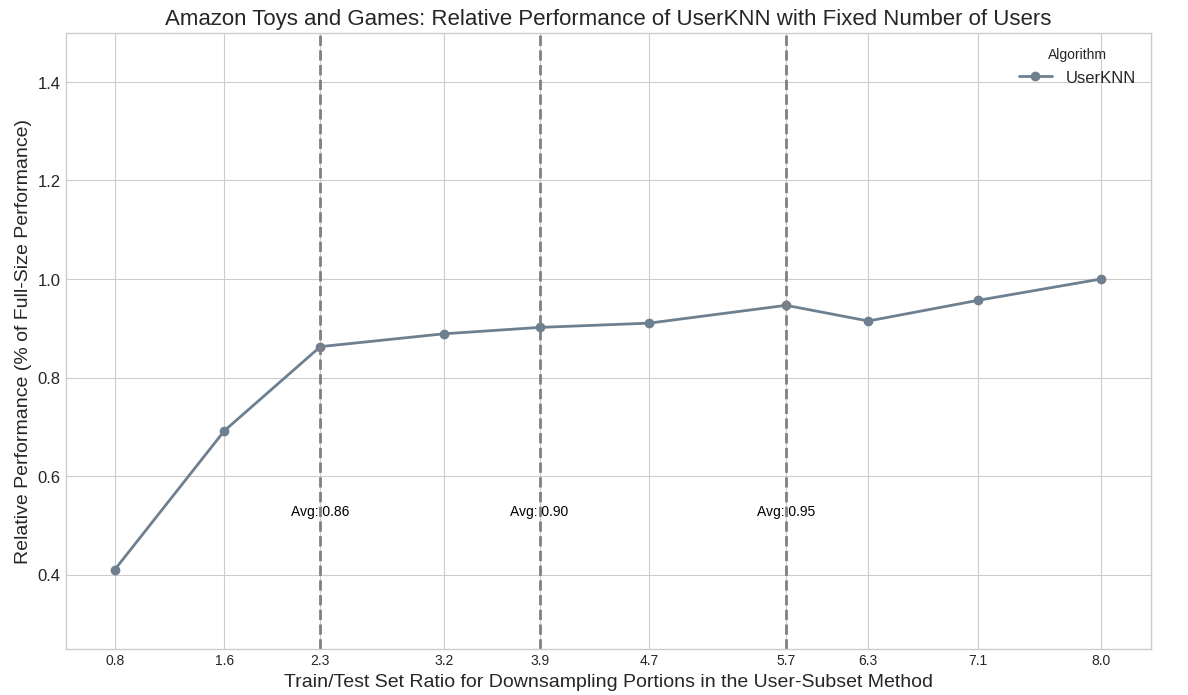}  
    \end{tabular}
    \caption{Relative Performance (\% of Full Dataset Baseline) of \textit{UserKNN} Algorithm on \textit{MovieLens 100K} and \textit{Amazon Toys and Games} with Fixed Number of Users Across Varying Train/Test Ratios.}
    \label{fig:Train/Test}
\end{figure}

However, when analyzing the impact of varying each user's train/test ratio while keeping the number of users involved fixed (Figure \ref{fig:Train/Test}), we observe trends consistent with our earlier observations: a downward trend for \textit{MovieLens} datasets and an upward trend for \textit{Amazon} datasets. This confirms that the variation in each user’s train/test ratio is the primary factor driving the contrasting behavior of the same group of algorithms across different datasets. This variation plays a critical role in explaining the unexpected performance patterns observed for some algorithms under the User-Subset approach (where obtained \textit{nDCG@10} score exceeds that of the full dataset at lower portions of the data). Ultimately, this factor accounts for the differences in algorithm behavior between the two downsampling approaches, emphasizing the influence of how data partitioning strategies impact performance. These findings align with prior research by Rocío et al., which emphasizes the significant influence of train/test splits on both absolute \textit{nDCG@10} scores and the relative performance of algorithms \cite{Canamares2020}.

Furthermore, the observed trends, whether upward or downward, appear to depend on both the dataset characteristics and the complexity of the algorithm. Our \textbf{assumption} is that, Simpler algorithms (e.g., \textit{Popular}, \textit{Bias}) or those with limited pattern discovery capabilities tend to struggle as the test set ratio decreases. With fewer interactions available for prediction, their chances of correct predictions diminish. These algorithms rely on straightforward patterns, such as popularity trends or biases, which do not improve significantly with a larger training set. Consequently, the expanded training set fails to offset the challenges posed by a smaller test set, leading to a consistent downward trend across both dense (\textit{MovieLens}) and sparse (\textit{Amazon}) datasets.

In contrast, algorithms which are better equipped to capture complex patterns in user-item interactions (e.g., \textit{UserKNN}, \textit{ItemKNN}, \textit{NeuMF}), exhibit dataset-dependent behaviors. In dense datasets like \textit{MovieLens}, where users have numerous interactions, these algorithms can learn sufficiently strong patterns even with smaller portions of training data. However, reducing the test set ratio for each user's interactions diminishes the opportunity to validate the learned patterns accurately. This reduction in the test set ratio appears to have a greater negative impact on performance than the incremental benefit gained from increasing the training set ratio, resulting in the observed downward trend.

Conversely, in sparse datasets like \textit{Amazon} and \textit{Gowalla}, where users generally have fewer interactions, increasing the training set ratio provides advanced algorithms with a larger pool of interactions to uncover meaningful patterns. The expanded training data offers a more comprehensive representation of user preferences, which outweighs the drawbacks of a smaller test set ratio, resulting in an upward performance trend. Sparse datasets, by nature, require more training data to support effective pattern discovery, making the increased training portion especially beneficial.

These observations suggest that dataset characteristics play a significant role in shaping the contrasting behaviors of algorithms under different downsampling scenarios. This aligns with findings by Adomavicius and Zhang, who highlighted the critical impact of data density and rating distribution on recommendation accuracy ~\cite{Adomavicius2012}. While these explanations offer a plausible basis for the observed trends, the precise mechanisms remain unclear. Further research is required to better understand the interplay between dataset properties and algorithm performance.

Based on the results and the details discussed regarding the downsampling strategies in this thesis, it is essential to emphasize that the two approaches, User-Based and User-Subset, are not directly comparable in terms of efficiency in maintaining algorithm performance. Their distinct configurations, focus areas, and the differing test sets they employ make a direct comparison impractical. Instead, the key aspect where they can be meaningfully compared is their impact on runtime and carbon footprint reduction. One of the primary motivations for introducing the User-Subset approach as an alternative to the widely used User-Based method was to analyze how reducing the dataset by decreasing the number of involved users affects runtime and carbon-equivalent reductions. In this aspect, the User-Subset approach demonstrates $\sim$14.5\% greater runtime efficiency on average across all algorithms, datasets, and downsampling portions compared to the User-Based method.

The User-Based approach remains a widely adopted and straightforward method for downsampling and dataset splitting \cite{meng2020exploring}. Its simplicity and intuitive setup make it the default choice in most research contexts. In contrast, the User-Subset approach, as introduced in this thesis, is not commonly used in similar studies. Its primary purpose here was to explore the effects of dataset reduction on runtime and carbon efficiency while maintaining the split setup defined in this thesis: allocating 10\% of total interactions for validation, 10\% for test, and varying portions for training set.

Although the User-Subset approach provides valuable insights, it introduces additional complexity. Specifically, it requires adjusting the train/test ratio for each user’s interactions and dynamically includes different users in the validation and test sets across downsampling portions. Although this setup is statistically consistent with the split methodology, the variability in user interaction patterns may influence the observed algorithm performance. These factors highlight that the User-Subset method is not without trade-offs.

Therefore, the User-Subset approach should be regarded as a case-study method designed to explore specific trade-offs and opportunities for analysis, particularly concerning runtime and carbon efficiency, rather than as a universally superior alternative to the User-Based approach.

Regarding \textbf{the effect of pruning}, the presented statistics indicate that the 30-core pruned versions of datasets generally lead to improved absolute algorithm performance. For the three \textit{MovieLens} datasets (\textit{100K}, \textit{1M}, \textit{10M}), the 30-core pruned versions, which eliminate 17\% of users compared to the 10-core versions, show an average improvement of $\sim$5.1\% and $\sim$8.9\% in \textit{nDCG@10} scores across all algorithms and portions for the User-Based and User-Subset approaches, respectively. Similarly, for the \textit{Gowalla} dataset, the 30-core pruning retains only users with 30 or more interactions, resulting in remarkable average score improvements of $\sim$205\% and $\sim$234\% in the User-Based and User-Subset approaches, respectively, despite eliminating 96\% of users compared to the 10-core version.

This behavior aligns with findings by Beel et al.~\cite{Beel2019}, who demonstrated that pruning datasets to exclude users with fewer interactions often leads to artificially inflated performance metrics because data points where algorithms typically underperform are removed. For example, their study highlighted that in pruned versions of the \textit{MovieLens} datasets, users with fewer than 20 ratings, representing 42\% of the user base but contributing only 5\% of the ratings, are excluded. Algorithms tend to perform worse for these low-interaction users, with \textit{RMSE} scores approximately 23\% higher compared to users with more interactions. These results suggest that excluding sparse users simplifies the modeling task and skews performance metrics favorably, a pattern also observed in the 30-core pruned datasets analyzed in this thesis.

When evaluating the ranking of algorithms, our experiments reveal only minor shifts in ranking between the 10-core and 30-core pruning versions in most cases, with differences in scores being minimal. However, in the heavily pruned \textit{Gowalla} dataset, algorithm rankings change more noticeably, indicating that the degree of pruning can influence the relative order of algorithms. This observation supports the hypothesis that substantial pruning can alter how algorithms perform relative to one another~\cite{Beel2019}.

Interestingly, the relative performance of algorithms, compared to their full-size scores, exhibits different trends depending on the downsampling approach. In the User-Based approach, algorithms trained on the 30-core pruned datasets show slightly lower averaged relative performance compared to the 10-core pruned datasets, with an average drop of around 2\%. Conversely, in the User-Subset approach, the 30-core pruned datasets show an average $\sim$4\% higher relative performance than their 10-core counterparts, highlighting an advantage in terms of energy efficiency-performance trade-offs.

While the primary focus of this project is not on the effects of pruning, the observed results underscore the importance of this factor and suggest that more focused and in-depth investigations into the implications of different pruning levels could provide valuable insights.

These findings and discussions naturally lead to addressing the central research question of this thesis:

RQ: \textit{Is it possible to achieve an acceptable trade-off between energy efficiency and performance in recommender algorithms by reducing dataset size?}

The findings of this study indicate that data reduction through downsampling can indeed contribute to energy efficiency by significantly reducing runtime and carbon emissions. As shown in Figure \ref{fig:Runtime}, this energy savings is closely correlated with the extent of data reduction, demonstrating the potential for improved efficiency in computationally intensive tasks.

However, the impact of data reduction on algorithm performance is highly scenario-dependent and varies with the downsampling approach. For example, in the User-Based downsampling method, data reduction typically results in a predictable decline in performance, as illustrated in Figure \ref{fig:Averaged-Normalized}. However, the extent of this performance drop varies across different groups of algorithms; some algorithms exhibit lower sensitivity to training data reduction, making them more suitable choices when energy efficiency is the primary focus. In contrast, the User-Subset approach introduced in this thesis demonstrates that, with a flexible and distinct design setup, it is possible to achieve metric scores that are comparable to or, in some cases, better than those achieved with the full dataset (strictly in terms of the absolute scores obtained), all while significantly reducing runtime and carbon footprint.

Ultimately, achieving a favorable trade-off between energy efficiency and performance depends on several factors: the complexity and nature of the algorithms (Figure \ref{fig:Confidence-Algorithm}), the characteristics of the datasets (Figure \ref{fig:Confidence-Dataset}), and the specifics of the downsampling setup. By carefully selecting and aligning these factors in scenarios where absolute peak performance is not the primary objective, and where flexibility in algorithm and dataset choice exists, researchers and practitioners can significantly improve energy efficiency with minimal and carefully optimized compromise to performance effectiveness.

\chapter{Conclusion}
This thesis investigates the potential of dataset downsampling as a strategy to achieve energy-efficient recommender systems while maintaining acceptable performance levels. By evaluating the effects of downsampling across diverse algorithms, datasets, and configurations, this research provides a comprehensive assessment of the trade-offs between energy efficiency and recommendation quality.

The results demonstrate that downsampling can effectively reduce the energy demands of recommender systems, leading to significant savings in runtime and carbon emissions. When considering 30\% up to 70\% downsampling portions, runtime reductions ranged from approximately 52\% up to 18\% (Figure \ref{fig:Runtime}), depending on the method and exact portion of downsampling, while carbon emissions were lowered by an estimated 17.66 up to 51.02 \textit{KgCO$_2$e} for training of a single algorithm on a single dataset. These findings highlight the value of data reduction in mitigating the environmental impact of AI-driven systems, particularly in large-scale applications.

In terms of performance, the study shows that recommender systems can retain a substantial portion of their effectiveness even with reduced training datasets. Relative performance ranged from $\sim$39\% to as high as $\sim$115\% of the full dataset scores averaged across algorithms and datasets (Figure \ref{fig:Averaged-Relative}), depending on the downsampling approach and portion used. This suggests that, in some cases, depending on the dataset reduction configuration and objectives, it is possible to achieve comparable or even better metric scores with smaller datasets, offering a practical pathway to optimize energy efficiency without incurring severe performance penalties.

The research also reveals that the effectiveness of downsampling strategies in achieving an optimal trade-off between energy efficiency and performance is influenced by the interaction of the characteristics of the data set and the algorithmic complexity (Figures \ref{fig:Confidence-Algorithm} and \ref{fig:Confidence-Dataset}). Sparse datasets, which inherently provide fewer interactions, typically result in lower absolute \textit{nDCG@10} scores compared to denser datasets. However, they could exhibit higher efficiency in data reduction under specific configurations of downsampling approaches and targeted data reductions. Similarly, simpler algorithms, while generally achieving lower overall performance compared to advanced algorithms capable of capturing intricate user-item patterns, generally demonstrated greater potential for achieving performance efficiency at reduced dataset portions due to their reduced sensitivity to training data size.

Finally, based on the findings of this thesis, we address our research question and conclude that downsampling can be a viable and effective strategy for balancing energy efficiency and performance in recommender systems, although its effectiveness remains highly context-dependent. Achieving a successful trade-off depends on factors such as the configuration of the downsampling approach, the characteristics of the datasets, and the complexity of the algorithms. Although absolute peak performance may occasionally require the full dataset, this study demonstrates that, for many practical applications, careful downsampling can achieve a favorable balance. This approach not only reduces computational resource requirements but also contributes to the development of sustainable AI systems.

In conclusion, this research supports the integration of downsampling techniques into the design and operation of recommender systems as a means of optimizing both performance and energy efficiency. By achieving meaningful reductions in resource usage with minimal impact on recommendation quality, downsampling offers a scalable and environmentally conscious solution to advance the field of recommender systems.

\chapter{Summary}
This thesis addresses the computational and environmental challenges posed by the growing size of datasets used in recommender systems. While larger datasets often enhance performance, they do so at the cost of increased energy consumption and carbon emissions. The central research question investigates whether reducing dataset size can balance energy efficiency and algorithmic performance without significant or with only minimal compromises. To explore this, the study utilized seven datasets, applying two distinct downsampling strategies, User-Based and User-Subset, along with two pruning levels (10-core and 30-core). A diverse set of 12 algorithms, ranging from basic to advanced, were evaluated using \textit{nDCG@10} as the performance metric. These configurations enabled a detailed analysis of the trade-offs between dataset size, algorithm performance, and energy efficiency, with a focus on identifying sustainable practices for recommender systems. Key findings of the research include:

\textbf{Overall Performance Trends}: The performance of the algorithms, measured by normalized \textit{nDCG@10}, consistently improved with larger training datasets in the User-Based approach. However, some algorithms displayed less sensitivity to increases in training data, suggesting they may perform relatively well even with smaller datasets.

In contrast, the User-Subset approach showed varied behavior. Some algorithms, particularly simpler or more basic ones, exhibited a downward trend in performance as the downsampling portion increased, yet still achieved metric scores exceeding those obtained with the full dataset at lower downsampling portions. For instance, as a representative case study at 50\% downsampling, the User-Based method achieved approximately 64\% of its full dataset \textit{nDCG@10} score, while the User-Subset approach, utilizing a different data reduction configuration, achieved approximately 110\%.

\textbf{Effect of Pruning}: The analysis of pruning levels showed that 30-core pruning improved absolute performance, with gains of $\sim$5.1\% to $\sim$8.9\% in dense datasets (\textit{MovieLens}) and up to $\sim$234\% in sparse datasets (\textit{Gowalla}). However, relative performance trends varied: 10-core pruning yielded higher efficiency under the User-Based method, while 30-core pruning excelled in the User-Subset approach. These results highlight the trade-offs between denser interaction networks and relative performance efficiency.

\textbf{Algorithmic Behavior}: Algorithms were categorized into two groups based on their performance trends. Group 1 algorithms, including \textit{ItemKNN} (both versions), \textit{UserKNN}, \textit{NeuMF}, \textit{SVD}, and \textit{NMF}, consistently benefited from larger training sets under both downsampling approaches, particularly under User-Based method, achieving higher \textit{nDCG@10} scores compared to the other group. In contrast, Group 2 algorithms namely \textit{Popular}, \textit{BiasedMF}, \textit{FunkSVD}, \textit{Popularity}, and \textit{Bias} exhibited higher relative performance under both approaches, as their \textit{nDCG@10} scores at reduced training set sizes were closer to their full-size dataset scores. These trends underscore the varying sensitivities of algorithms to dataset size and downsampling strategies.

\textbf{Dataset Characteristics}: Performance trends varied across the dataset groups of \textit{MovieLens}, \textit{Amazon}, and \textit{Gowalla}. On average, \textit{MovieLens} datasets achieved higher \textit{nDCG@10} scores and showed better relative performance (efficiency) in the User-Subset method. Conversely, \textit{Amazon} and \textit{Gowalla} datasets demonstrated better relative performance at higher downsampling portions (above 40\%) under the User-Based approach. These findings highlight the impact of characteristics of the dataset on algorithmic behavior.

\textbf{Environmental Benefits}: Downsampling significantly reduced runtime and carbon emissions. For instance, at 30\% downsampling, depending on the chosen downsampling approach, runtime reductions ranged from $\sim$36\% to $\sim$52\% compared to full-size datasets, with corresponding carbon savings of $\sim$35.32 to $\sim$51.02 \textit{KgCO$_2$e} per algorithm per dataset. Notably, the User-Subset approach consistently outperformed the User-Based method in reducing runtime and energy consumption.

\textbf{Impact of Downsampling Strategies}: The User-Based approach, characterized by its commonly used configuration where only the interactions of the same consistent users are downsampled, provided predictable results, aligning with the traditional expectation that larger training set yield better outcomes. In contrast, the User-Subset method revealed unexpected trends, with certain algorithms benefiting from reduced dataset sizes in terms of achieved absolute scores compared to the full dataset score. These trends can be attributed to differences in configurations, such as variations in each user's train/test ratios and user selection mechanisms.

Overall, this thesis demonstrates the feasibility of optimizing dataset sizes to achieve sustainable AI practices. However, the ability to achieve a favorable trade-off between energy efficiency and performance depends on multiple factors, including algorithm complexity, dataset characteristics, and downsampling configurations. The findings suggest that smaller datasets are capable of reducing runtime and carbon footprint without necessarily sacrificing performance significantly in many cases. As recommender systems are increasingly used in large-scale applications, adopting such optimization strategies can have a meaningful impact on reducing the environmental footprint of AI technologies. This research contributes to the emerging field of Green Recommender Systems by offering practical guidelines to balance performance and sustainability.

\chapter{Future Work and Limitations}
This research provides a comprehensive analysis of the trade-offs between energy efficiency and recommendation quality through dataset downsampling. However, several limitations and areas for future exploration remain that could enhance the applicability and impact of this work.

\textbf{1. Scope of Algorithms}

While this thesis evaluates a diverse range of 12 algorithms across multiple libraries (LensKit, RecPack, and RecBole), the selection is not exhaustive. Most of the algorithms examined are traditional recommender systems, with \textit{NeuMF} being the only deep learning model analyzed. This limited inclusion restricts insights into how modern, cutting-edge models respond to downsampling. Future research should include more state-of-the-art deep learning algorithms, particularly those based on graph neural networks and transformer-based architectures. Examining these advanced models would provide a deeper understanding of the impact of downsampling strategies on algorithm performance and efficiency, offering a more comprehensive evaluation of contemporary recommender systems.

\textbf{2. Dataset Characteristics and Core Pruning}

The datasets analyzed in this study span multiple domains and levels of sparsity. However, highly sparse datasets, particularly from the \textit{Amazon} collection, were excluded under certain pruning conditions (e.g., 30-core pruning) due to insufficient interactions after pruning. This limits the generalizability of findings related to pruning effects, especially for datasets with extreme sparsity or unique interaction patterns. Future research should incorporate a broader spectrum of datasets, including highly sparse datasets from both explicit and implicit feedback types, that remain viable under different core pruning levels, to validate findings across diverse contexts. Additionally, while core pruning was a secondary focus in this thesis, its notable effects on algorithm rankings and performance trends suggest that dedicated investigations into varying pruning levels could yield valuable preprocessing guidelines for recommender systems.

\textbf{3. Downsampling Configuration and Train/Test Ratio Dynamics}

The User-Subset approach introduced in this study demonstrated effectiveness but also introduced complexity due to its dynamic train/test ratio and selective user inclusion. Moreover, the approach involves varying groups of users in the validation and test sets across different downsampling portions, which, while statistically consistent in maintaining fixed sizes for these sets, may influence observed performance. This complexity complicates direct comparisons with simpler methods and may limit its applicability in scenarios where simplicity and standardization is preferred. Future work could focus on developing more straightforward and commonly used downsampling approaches or innovative techniques that address these limitations, maintaining efficiency while offering simpler and more practical configurations. Furthermore, in this study, the experiments were conducted using a fixed Train/Validation/Test split ratio of 80\%-10\%-10\%. Exploring alternative split configurations (e.g., 60\%-20\%-20\%), could provide further insights into the efficiency and effectiveness of different train/test ratios in downsampling scenarios Moreover, further exploration is needed to understand the dynamic interaction between train/test ratios and algorithm performance across datasets. For instance, advanced algorithms exhibited improved performance on sparse datasets (e.g., \textit{Amazon}, \textit{Gowalla}) with increasing training portions, while the same group of algorithms showed decreased performance in dense datasets (e.g., \textit{MovieLens}). Investigating these contrasting behaviors would help optimize downsampling configurations to maximize efficiency and effectiveness for various algorithm-dataset combinations.

\textbf{4. Environmental Impact Assumptions and Broader Metrics}

This study approximates carbon emissions and energy savings based on assumptions such as linear correlations between runtime, energy consumption, and \textit{CO$_2$e} emissions. Although useful for obtaining an initial understanding, these estimates may vary significantly depending on real-world factors such as hardware, software configurations, and energy sources. Future research should consider additional sustainability metrics, such as the impact of downsampling on hardware and computational resource demands (e.g., RAM and GPU usage). Expanding sustainability metrics beyond carbon emissions and runtime would provide a more comprehensive view of the environmental implications of downsampling strategies.

\textbf{5. Limited Evaluation Metrics}

Performance evaluation in this research is based solely on the \textit{nDCG@10} metric, which measures the quality of ranking of the top 10 recommendations. While \textit{nDCG@10} is relevant for ranking-focused tasks, it does not encompass other important aspects of recommender systems, such as precision, recall, diversity, and novelty. Future studies should incorporate a broader range of evaluation metrics to provide a more holistic assessment of performance. This would allow researchers to better understand the trade-offs introduced by downsampling, particularly its broader impact on the quality and utility of recommendations.

The findings of this research contribute significantly to the emerging field of Green Recommender Systems, highlighting the potential of downsampling as a strategy to balance sustainability and effectiveness. However, addressing these limitations and pursuing the proposed future directions will help further advance the development of efficient, scalable, and environmentally responsible recommender systems.

\newpage 

\addcontentsline{toc}{chapter}{Bibliography} 
\bibliographystyle{plainurl} 
\bibliography{reference} 

\chapter*{Declaration of Authorship}
\pagenumbering{roman} 
\setcounter{page}{8} 
\addcontentsline{toc}{chapter}{Declaration of Authorship} 

\noindent
I hereby confirm that this thesis and the work presented in it is entirely my own. Where I have consulted the work of others, this is always clearly stated. All statements taken literally from other writings or referred to by analogy are marked, and the source is always given. This paper has not yet been submitted to another examination office, either in the same or similar form.\\[3cm]
\noindent

\begin{minipage}[t]{0.4\textwidth}
    \textbf{Place and Date:}\\
    Siegen, 31.01.2025\\[2cm]
\end{minipage}
\hfill
\begin{minipage}[t]{0.4\textwidth}
    \textbf{Name:}\\[0.2cm]
    Ardalan Arabzadeh    

\end{minipage}

\end{document}